\documentclass[conference]{IEEEtran}
\IEEEoverridecommandlockouts
\usepackage{soul}
\usepackage{enumitem}
\usepackage[utf8]{inputenc}
\usepackage{epsfig, subfigure, amsmath, amssymb}
\usepackage{textcomp}
\usepackage{url}

\usepackage{amsmath,amssymb,amsfonts}
\usepackage[ruled,vlined,linesnumbered]{algorithm2e}
\usepackage{algorithmic}
\usepackage{graphicx}
\usepackage{textcomp}
\usepackage{multirow}
\usepackage{xcolor}
\usepackage{paralist}

\definecolor{red}{rgb}{1,0,0}
\definecolor{orange}{rgb}{1,0.64,0.06}
\definecolor{emerald}{rgb}{0.00784313725 ,0.54117647058,0.05882352941}

\def\BibTeX{{\rm B\kern-.05em{\sc i\kern-.025em b}\kern-.08em
    T\kern-.1667em\lower.7ex\hbox{E}\kern-.125emX}}

\author{
\IEEEauthorblockN{Nur Imtiazul Haque\IEEEauthorrefmark{1}, Maurice Ngouen\IEEEauthorrefmark{1}, Mohammad Ashiqur Rahman\IEEEauthorrefmark{1}, Selcuk Uluagac\IEEEauthorrefmark{2}, and Laurent Njilla\IEEEauthorrefmark{3}}
\IEEEauthorblockA{\IEEEauthorrefmark{1}Analytics for Cyber Defense (ACyD) Lab, Florida International University, USA\\
\IEEEauthorrefmark{2}Cyber-Physical Systems Laboratory (CPSLab), Florida International University, USA\\
 \IEEEauthorrefmark{3}US Air Force Research Laboratory (AFRL), USA\\
	\IEEEauthorrefmark{1}\{nhaqu004, mngou002, marahman\}@fiu.edu, \IEEEauthorrefmark{2}suluagac@fiu.edu, \IEEEauthorrefmark{3}laurent.njilla@us.af.mil
	}
}

\begin{document}

\title{SHATTER: Control and Defense-Aware Attack Analytics for Activity-Driven Smart Home Systems}

\maketitle
\pagestyle{plain}

\begin{abstract}
Modern smart home control systems utilize real-time occupancy and activity monitoring to ensure control efficiency, occupants' comfort, and optimal energy consumption. Moreover, adopting machine learning-based anomaly detection models (ADMs) enhances security and reliability. However, sufficient system knowledge allows adversaries/attackers to alter sensor measurements through stealthy false data injection (FDI) attacks. Although ADMs limit attack scopes, the availability of information like occupants' location, conducted activities, and alteration capability of smart appliances increase the attack surface. Therefore, performing an attack space analysis of modern home control systems is crucial to design robust defense solutions. However, state-of-the-art analyzers do not consider contemporary control and defense solutions and generate trivial attack vectors. To address this, we propose a control and defense-aware novel attack analysis framework for a modern smart home control system, efficiently extracting ADM rules. We verify and validate our framework using a state-of-the-art dataset and a prototype testbed.
\end{abstract}


%
\section{Introduction}
\label{sec:introduction}
Contemporary home control systems use enormous remotely accessible and controllable internet-connected smart devices to ensure energy efficiency and occupant\'s comfort. The adoption of smart devices is increasingly growing due to their affordability, 
accuracy, interoperability, productivity, cost reduction, and so on. Smart home control systems are currently assisted with voice-controlled smart devices (e.g., turning on the bedroom light through a smartphone voice assistant) or self-learned automated closed-loop controllers (e.g., smart cooling controller self-adjusted based on the homeowners' schedule). The prevalence of occupancy sensors and tracking devices (e.g., through smartwatches or RFID sensors) accounts for improved accuracy and efficiency of the control systems through real-time occupants' location and activity identification.

Unfortunately, the widespread use of the internet of things (IoT) network in smart devices has left smart home control systems highly susceptible to multiple cyberattacks. Such devices possess restricted security capabilities, leaving them vulnerable to constantly evolving and sophisticated attacks due to their open network communication. Hence, millions of IoT devices are currently functioning without adequate security protection~\cite{eskandari2020passban}. Since smart homes/buildings are susceptible to several well-known attacks such as ransomware, distributed denial of service (DDoS), and data manipulation, it is crucial to investigate the vulnerability of the heating, ventilation, and air conditioning (HVAC) system, which is a critical component of a home. Our security analysis considers false data injection (FDI) attacks on demand-controlled HVAC (DCHVAC) systems. We consider a sophisticated attacker having malicious intent to maximize the overall energy consumption. The attack motivation could be sabotage/rivalry/personal vendetta that projects financial loss to the home dwellers. While FDI attacks on smart homes are considered to be in the conceptual phase, instances of such attacks have been reported, as demonstrated by an attacker who boasted publicly of increasing a home's temperature by 20$^{\circ}$~F.~\cite{newskysecurity2022}.

The attack space analysis of a smart home control system is an active research area. In one of our existing works, we analyzed FDI attacks on a American Society of Heating, Refrigerating, and Air-Conditioning Engineers (ASHRAE)-based DCHVAC system (i.e., optimally mixes return and fresh air to meet energy efficiency and occupant's comfort) in the smart building context~\cite{haque2021biota}. A limited set of verification rules like maximum capacity of the zones, IAQ measurement consistencies, occupants count consistencies throughout the zones with the entrance count, etc., were considered in BIoTA for assessing the attack space of the home/ building control system. However, most modern smart home/building control systems are quite different than the assumption made in the existing works. Modern smart home control systems often use machine learning (ML)-based anomaly detection model (ADM) for identifying measurement inconsistencies, smart appliances control through voice assistants (through dedicated and other IoT device controllers), and occupants' activity monitoring tool. The ML-based ADM has already been adopted in industrial automation. e.g., BuildingIQ offers an intelligent energy management system that includes occupancy sensors and can adjust HVAC settings based on occupancy and building usage patterns~\cite{buildingiq2023}. Although not implemented in the industrial application, the activity recognition-based DCHVAC system has also been adopted in research facilities like KTH Live-In Lab, CASAS, ARAS testbeds~\cite{alemdar2013aras, kth2020, cook2012casas}. Hence, existing regulation-based approaches~\cite{stellios2021assessing, bakhshi2018industrial, akatyev2019evidence}, formal security analyses~\cite{mohsin2016iotsat, mohsin2017iotchecker}, ML-based approaches~\cite{ souri2020formal}, and ML-model verification~\cite{katz2017reluplex, dutta2017output, katz2019marabou} tools are inapplicable in such ADM-based smart home contexts since the ADMs learn the pattern of occupants' behavior, which makes the attacks considered in the existing works unstealthy. We propose \textbf{S}mart \textbf{H}ome \textbf{A}nalytics for \textbf{T}hreats \textbf{T}argeting \textbf{E}nergy \textbf{R}outine (SHATTER) framework that identifies critical threats of smart home control systems with ML-based ADM and activity identification modules. While the ADM limits the attack scope of SHATTER-identified attacks, the appliance-triggering attack utilizing the activity identification module increases the attack impact. Our evaluation shows that ML-based ADM reduces the attack impact by 50\% while leveraging the activity identification modules; an attacker can increase the attack impact by 20\% as compared to the state-of-the-art (i.e., BIoTA) framework. For formally modeling the ML-based ADM, we use a convex hull algorithm~\cite{barber1996quickhull}, where the constraint acquisition from the ML models is inspired by the SHChecker framework~\cite{haquenovel}. A satisfiability modulo theories (SMT)-based solver is used to identify optimal attack paths to launch stealthy FDI attacks in the considered smart home control system. We verify our proposed framework with two houses of state-of-the-art dataset naming Activity Recognition with Ambient Sensing (ARAS)~\cite{alemdar2013aras} and our built prototype testbed.
%
In summary, our contributions are as follows:
\begin{compactitem}
    \item We formally model a smart home HVAC control system with ML-based ADM and activity recognition module using first-order predicate logic by extracting constraints from the component models to analyze the system.
    \item We develop a threat analysis framework (SHATTER) to identify potential attack vectors in the smart home control system by formally modeling FDI attacks with variable attack attributes. 
    \item We conduct experiments with our formal threat analysis framework on state-of-the-art datasets and a real prototype testbed to identify critical attack vectors and evaluate the tool's scalability in analyzing the attack vectors.
\end{compactitem}
All implementation and evaluation results are reproducible with the source code on GitHub~\cite{shatter2022}.
The rest of the paper is organized as follows: 
we provide an overview of the considered smart home system and its components in Section~\ref{sec:control-system}. We provide a formal description of the problem domain and considered the attack model in Section~\ref{sec:problem-definition}.  
In the following section, we present the technical details of the proposed SHATTER framework. We provide case studies to give insights about our proposed framework's working principle and capabilities in Section~\ref{sec:case_studies}.
Then, we show the validation of the SHATTER framework with a real prototype testbed. 
We evaluate SHATTER using state-of-the-art datasets in Section~\ref{sec:evaluation}. A comprehensive literature review is presented in Section~\ref{sec:related-works}. We conclude the paper in Section~\ref{sec:conclusion}.

\section{Smart Home Control System}
\label{sec:control-system}

\begin{figure}[t]
\centering
\includegraphics[width = 0.96\columnwidth]{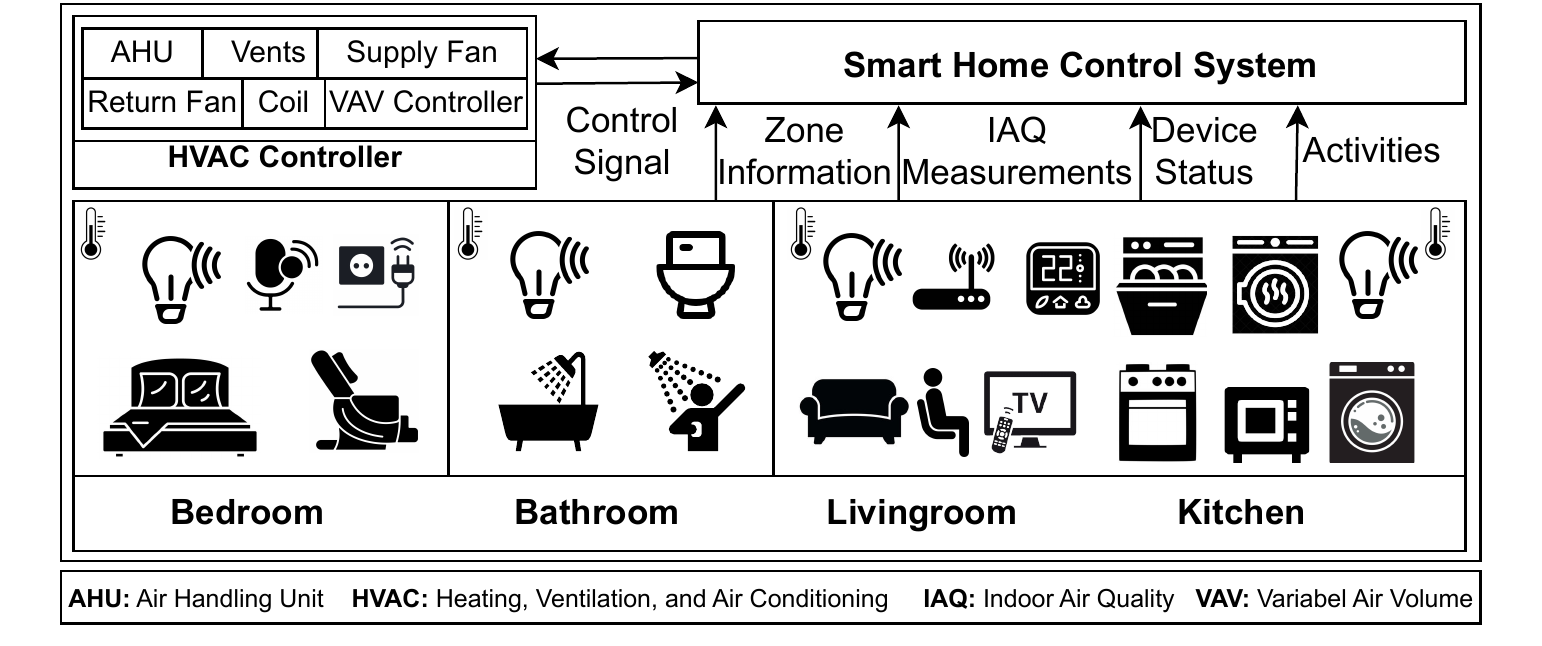}
\vspace{-10pt}
\caption{Smart home system with HVAC controller.}
\label{fig:system-overview}
\vspace{-10pt}
\end{figure}
\begin{figure}[t]
\centering
\includegraphics[width = 0.35\textwidth]{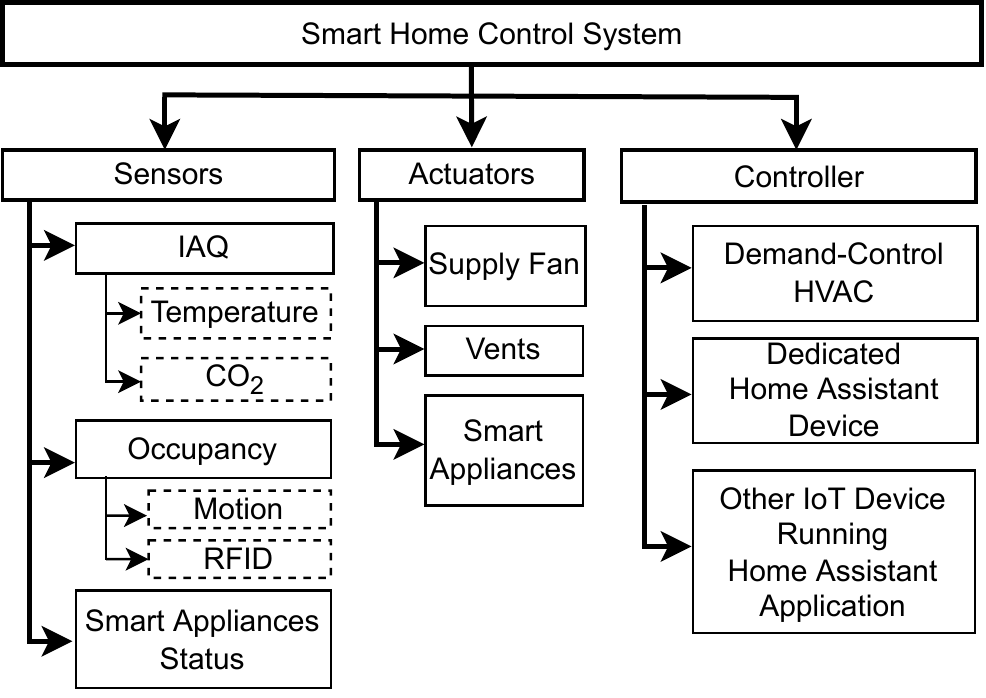}
\centering
\vspace{-9pt}
\caption{Components of smart home control systems.}
\label{fig:system-components}
\vspace{-5pt}
\end{figure}

We present a comprehensive but simplistic overview of the smart home control system considered in this work considering a DCHVAC system that can supply the optimal air to meet occupants' comfort and energy efficiency. Our considered smart home control system can track/locate the occupants in different zones and their conducted activities that allow predicting/estimating the IAQ (i.e., temperature and air pollutants) and hence the cooling/heating/ventilation demand. The close approximation of demand enables the calculation of optimal actuation of the control system. Figure~\ref{fig:system-overview} shows the architecture of the considered home control system, where the system acquires various IoT-based sensor information to estimate the smart home state (i.e., occupants' location, activity, and appliance status). The HVAC controller, which is the core controller of our considered control system, generates the optimal control signal to actuate the supply fan, return fan, and vents. We consider that all the appliances in the home are smart IoT devices and can be accessed, triggered, or actuated by the dedicated device or application-controlled voice assistants. The status of the smart appliances can be identified by the sensor installed on the appliances or the appliance control applications. Figure~\ref{fig:system-components} shows the hierarchy of the control system's different components. 


\begin{figure}[!t]
\vspace{-6pt}
    \begin{center}
         \subfigure[]
        {
        \label{subfig:control_cost_comparison_house_A}
            \includegraphics[width=0.48\columnwidth]{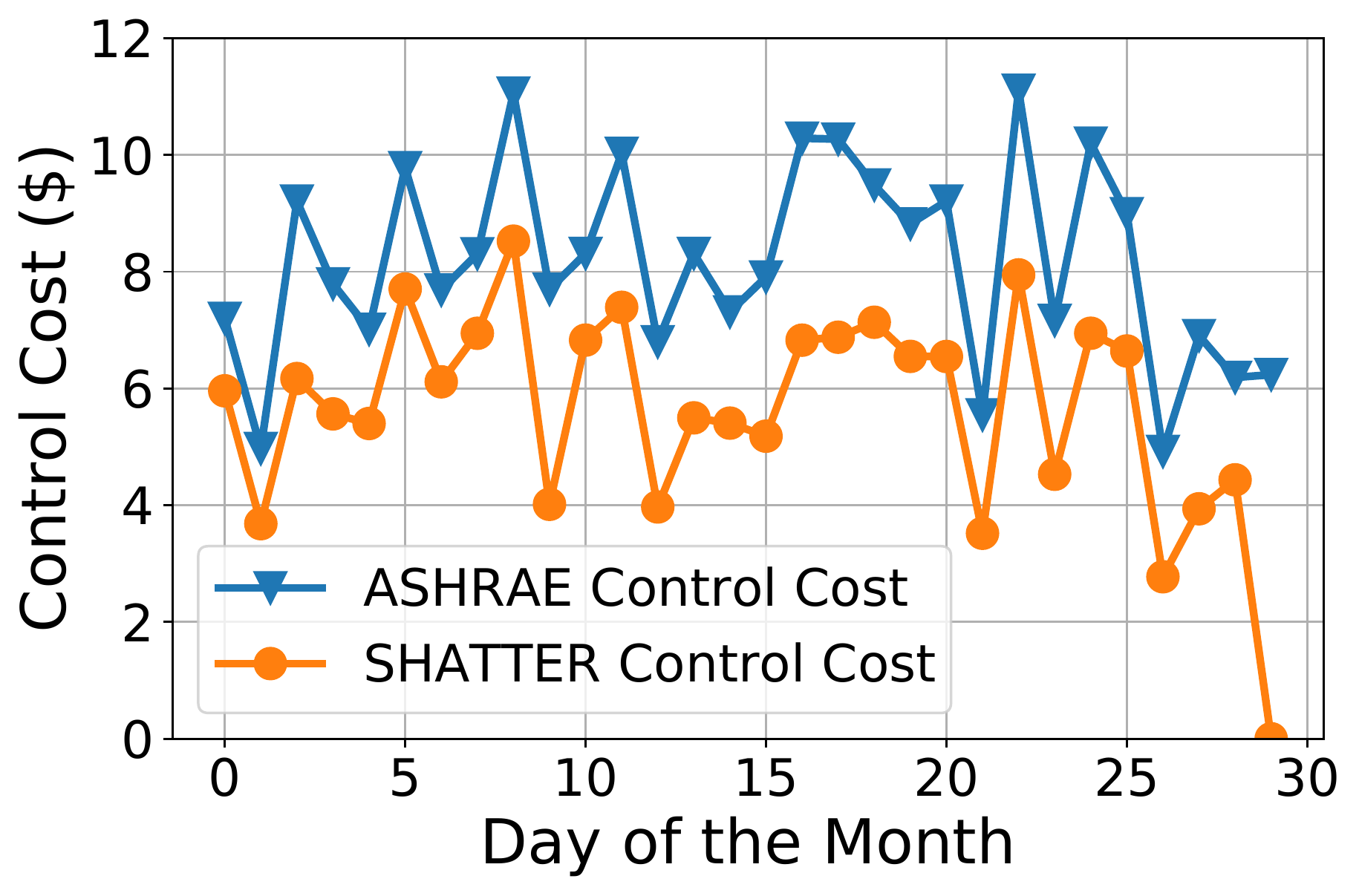}
        }\hspace{-9pt}
        \subfigure[]
        {
        \label{subfig:control_cost_comparison_house_B}
            \includegraphics[width=0.48\columnwidth]{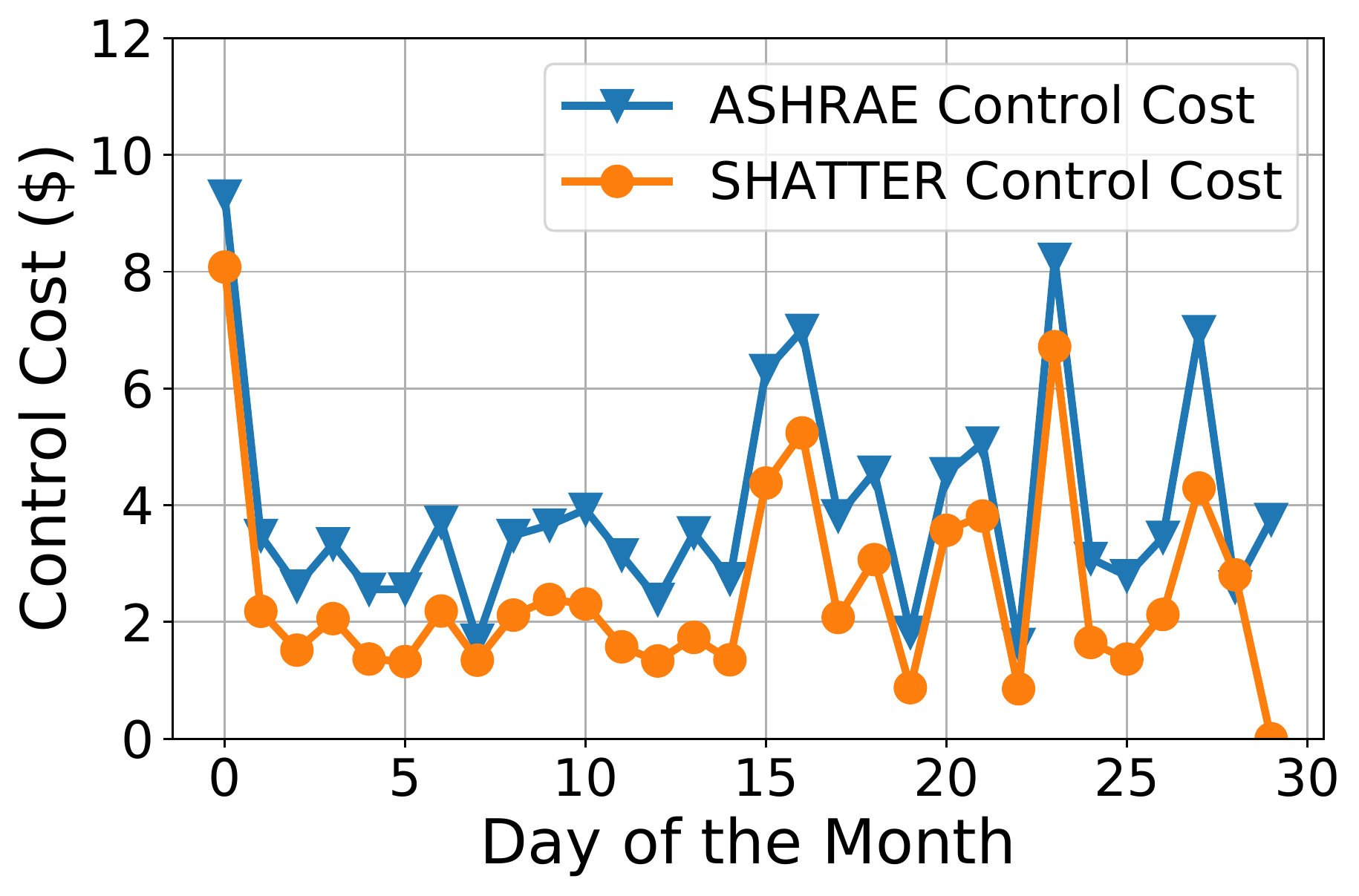}
        }     
    \end{center}
    \vspace{-12pt}
    \caption{Comparison between ASHRAE and proposed Control Cost (\$) for (a) ARAS House-A (b) ARAS House-B.}
    \label{fig:control_cost_comparison}
    \vspace{-17pt}
\end{figure}
Although we adopt the DCHVAC controller based on ASHRAE standards, some variations in our considered controller made it more efficient. Figure~\ref{fig:control_cost_comparison} shows the control cost comparison (the proposed DCHVAC controller is 48.2\% efficient for ARAS house-A, while 53.35\% efficient for house-B). It is to be noted that the purpose of this work is not to develop an efficient controller. However, the adoption of a sophisticated controller helps us identify critical attack vectors. The efficiency of the proposed controller is due to the following 3 reasons.
\begin{compactenum}[(1)]
    \item \textbf{Activity-Based Actuation} We consider an occupant activity-based DCHVAC controller, unlike the ASHRAE-based control model, which considers an average change in IAQ by the occupants. However, research by Persily et al.~\cite{persily2017carbon} shows that the level of physical activities of the occupants impacts the metabolic rate, in turn, the IAQ of the home. 
    \item \textbf{Activity-Appliance Relationship} The ASHRAE standard considers an average load (i.e., appliances) for the HVAC control system estimated by studying historical data. However, the estimated load is not good for meeting instantaneous demand. For instance, a person studying in the living room does not interact with any appliances and the control system with average load modeling will supply more air, thus will create discomfort for that person. Hence unlike BIoTA, we relate the appliances with conducted activity (i.e., appliance accessing information is used for activity recognition).
    \item \textbf{Occupants tracking} Another factor that contributed to developing an efficient controller is that the considered control system is continuously tracking zone-wise occupants through RFID sensors. Persily et al.~\cite{persily2017carbon} also identified that the occupant demographics influence the heat and pollutant generation in the zones. For instance, a middle-aged man generates twice as much air pollutants compared to an infant.
\end{compactenum}
%

The considered controller integrates an ML-based ADM for detecting sensor measurement inconsistencies, which is detailed in the following section.

\section{Problem Definition and Attack Model}
\label{sec:problem-definition}
This section provides a formal definition of the assumed home control system and a summary of the attack model. 
\subsection{Problem Definition}
We consider a smart home, $\mathbb{H}$ with smart sensors $\mathcal{S}$ and actuators $\mathcal{A}$, which are triggered by a control system $\mathbb{C}$. Both automated (e.g., HVAC controller) and manual controllers (e.g., smartphone sending voice commands to trigger a microwave in the kitchen) are part of $\mathbb{C}$. Different activities $\mathcal{D}$ of occupants, $\mathcal{O}$ residing in different zones, $\mathcal{Z}$ of $\mathbb{H}$ are constantly monitored through some $\mathcal{S}$ (e.g., RFID, photocell, contact, sonar distance sensors). The use of RFID sensor devices allows tracking the specific occupant/s residing in different $\mathcal{Z}$. The instantaneous activity information of $\mathcal{O}$ at different $\mathcal{Z}$ helps build a more energy-efficient HVAC controller (i.e., a component of $\mathcal{C}$) since different human activity correlates to different metabolic rates that directly control the IAQ of the zones. Other than the HVAC controller, our problem scope considers a smart home automation controller, which triggers smart devices (e.g., smart lights, smart kitchen utensils) throughout different $\mathcal{Z}$ using dedicated home assistant devices (e.g., Amazon Alexa) or other IoT device  (e.g., smartphone) controller applications. 

An ML-based (combined with some verification rule) ADM $\mathbb{E}$ checks the measurement consistencies at different timestamps. However, with the knowledge of $\mathcal{C}$ and $\mathbb{E}$, an attacker can still launch a stealthy FDI attack through intelligently crafting different $\mathcal{S}$. Suppose, there is only one occupant in $\mathbb{H}$, and he/she is staying in the bedroom zone. The occupant doing some chores will be alarmed if the washer or dryer is turned on through adversarial attempts, although unwanted turning on of the oven or microwave in the kitchen zone will be unnoticeable to the occupant. However, if the occupant is sleeping deeply and the bedroom door is closed, she will most likely be unaware of the adversarial activation of the washer or dryer. Hence, we consider an occupant activity model that learns the temporal behavioral and activity patterns. For example, if the occupant enters the bathroom at 2.00 pm, he/she is taking a shower for 20 minutes to 30 minutes, or if the occupant goes into the bedroom zone at 10 pm, he/she sleeps for 6 to 8 hours.
\begin{figure}[!t]
    \begin{center}
         \subfigure[]
        {
        \label{subfig:cluster_visualization_dbscan}
            \includegraphics[width=0.48\columnwidth]{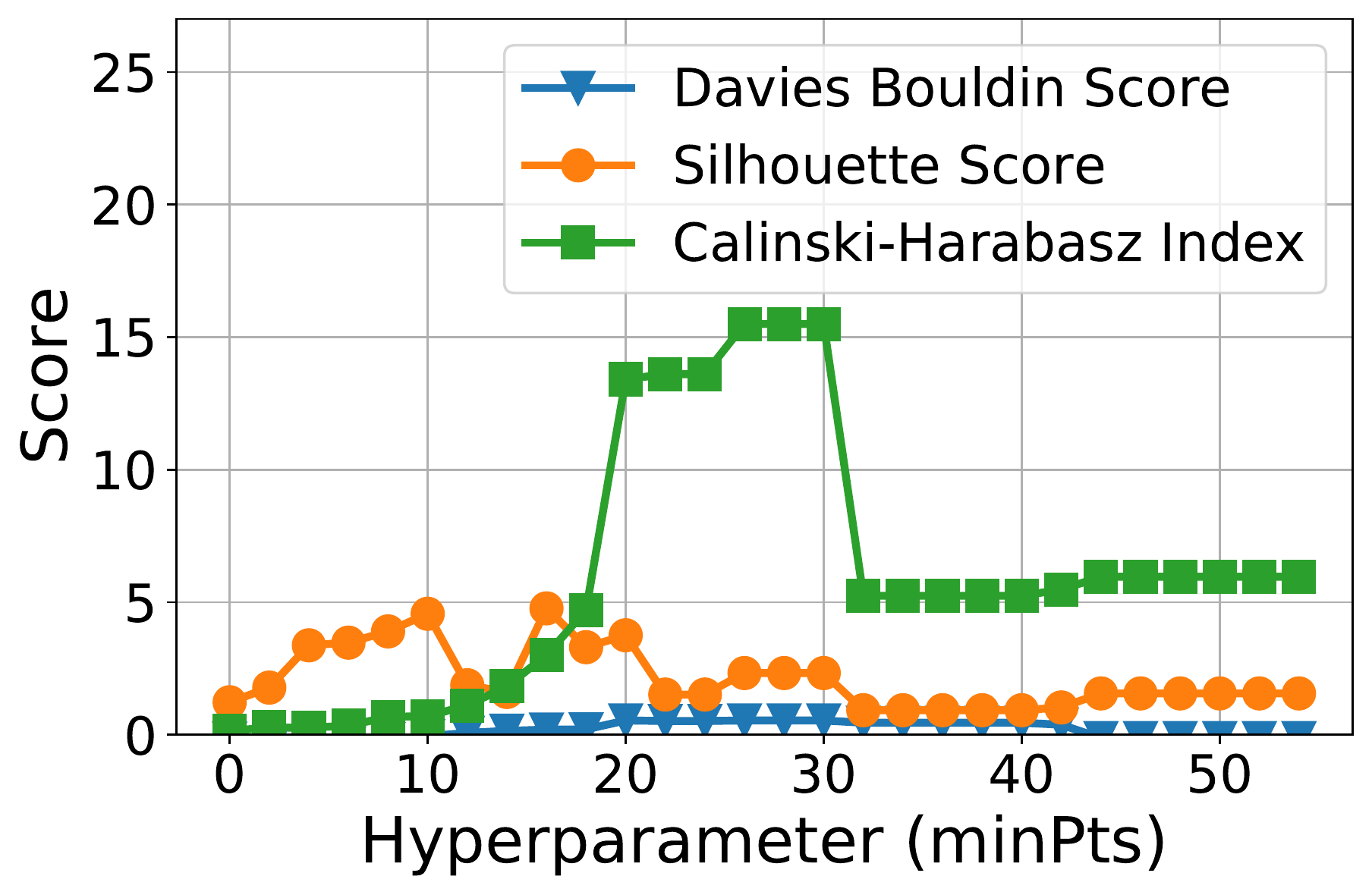}
        }\hspace{-9pt}
        \subfigure[]
        {
        \label{subfig:cluster_visualization_kmeans}
            \includegraphics[width=0.48\columnwidth]{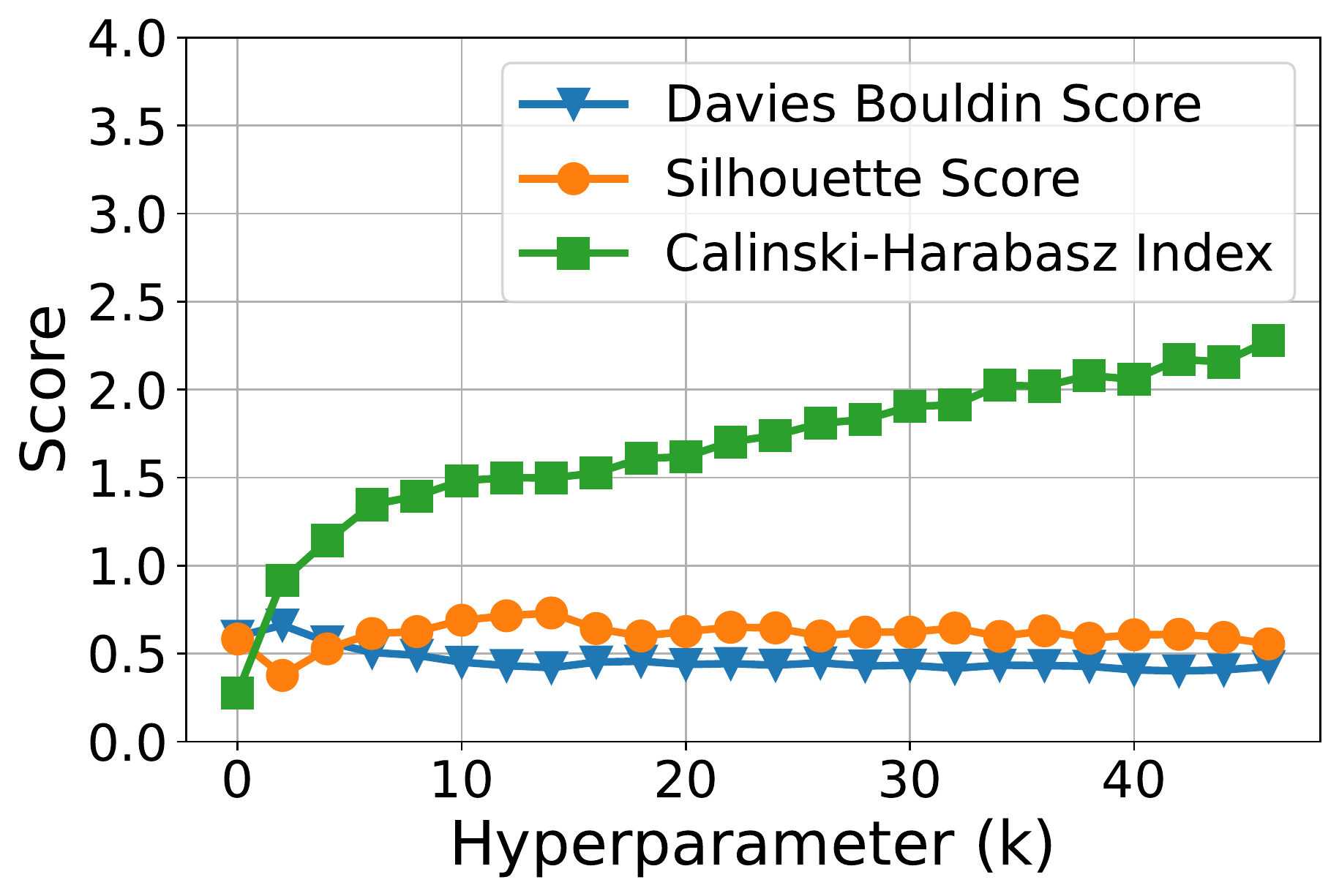}
        }     
    \end{center}
    \vspace{-14pt}\caption{Hyperparameter tuning of (a) DBSCAN and (b) K-Means clustering-based ADM for ARAS HAO1 dataset.}
    \label{fig:cluster_visualization_}
    \vspace{-12pt}
\end{figure}

\begin{figure}[!t]
    \begin{center}
         \subfigure[]
        {
        \label{subfig:dbscan_adm_progressive_performance}
            \includegraphics[width=0.46\columnwidth]{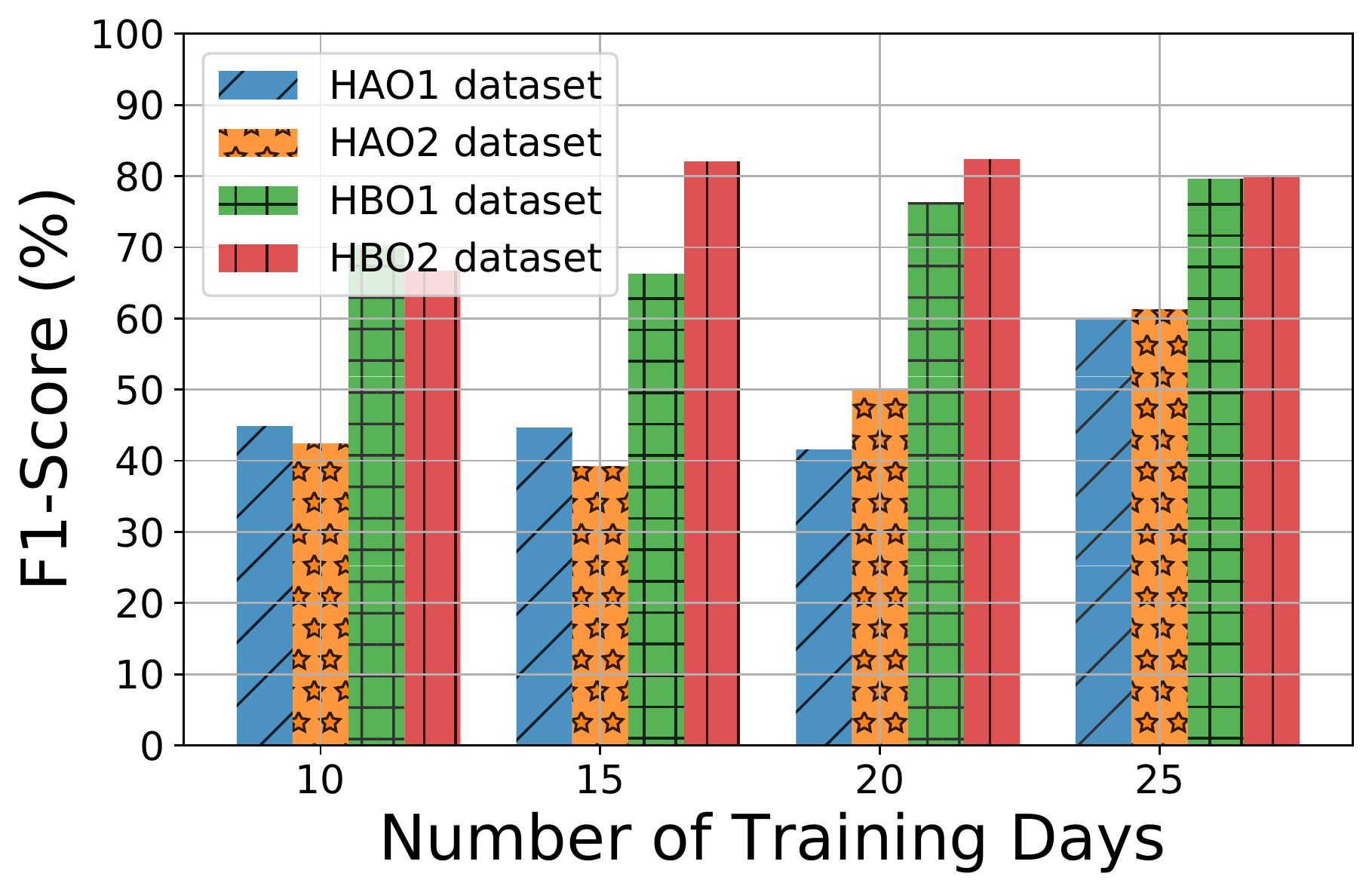}
        }
        \subfigure[]
        {
        \label{subfig:kmeans_adm_progressive_performance}
            \includegraphics[width=0.46\columnwidth]{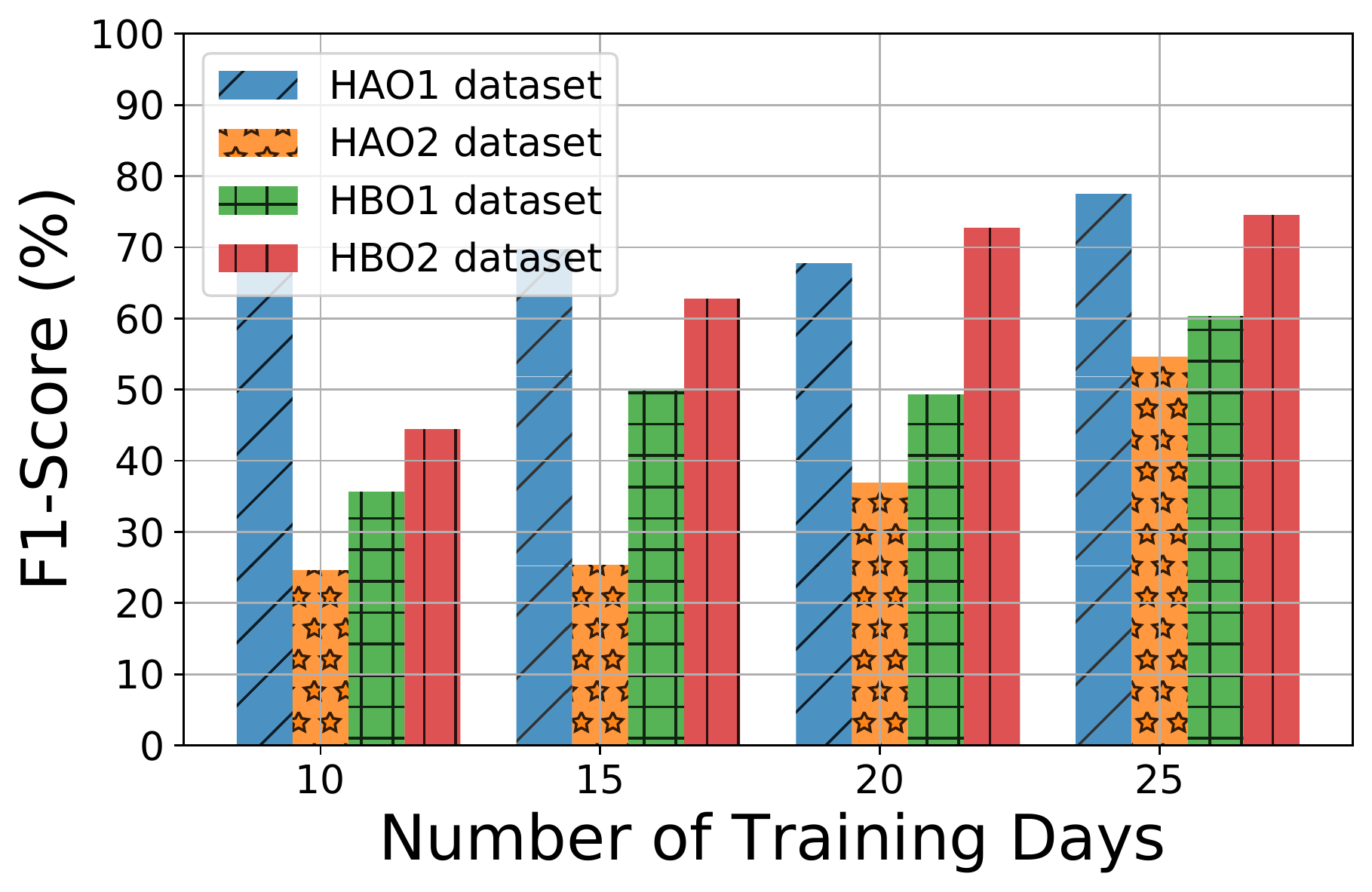}
        }     
    \end{center}
    \vspace{-5pt}
    \caption{Progressive incremental performance visualization for (a) DBSCAN and (b) K-Means clustering-based ADM based on F1-score.}
    \label{fig:anomaly_detection_progressive_performance}
    \vspace{-10pt}
\end{figure}

\begin{figure}[!t]
    \begin{center}
         \subfigure[]
        {
        \label{subfig:cluster_visualization_dbscan}
            \includegraphics[width=0.48\columnwidth]{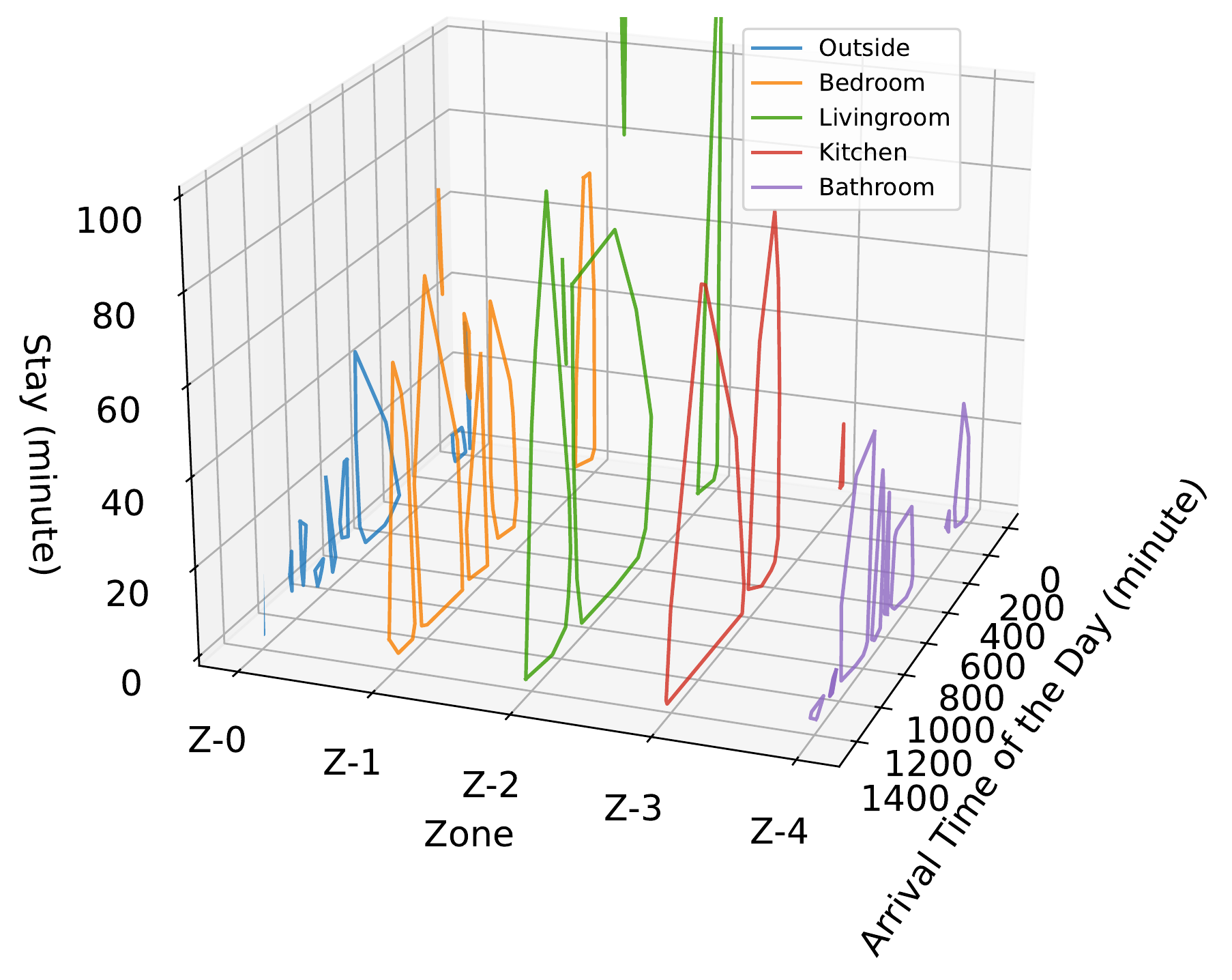}
        }\hspace{-9pt}
        \subfigure[]
        {
        \label{subfig:cluster_visualization_kmeans}
            \includegraphics[width=0.48\columnwidth]{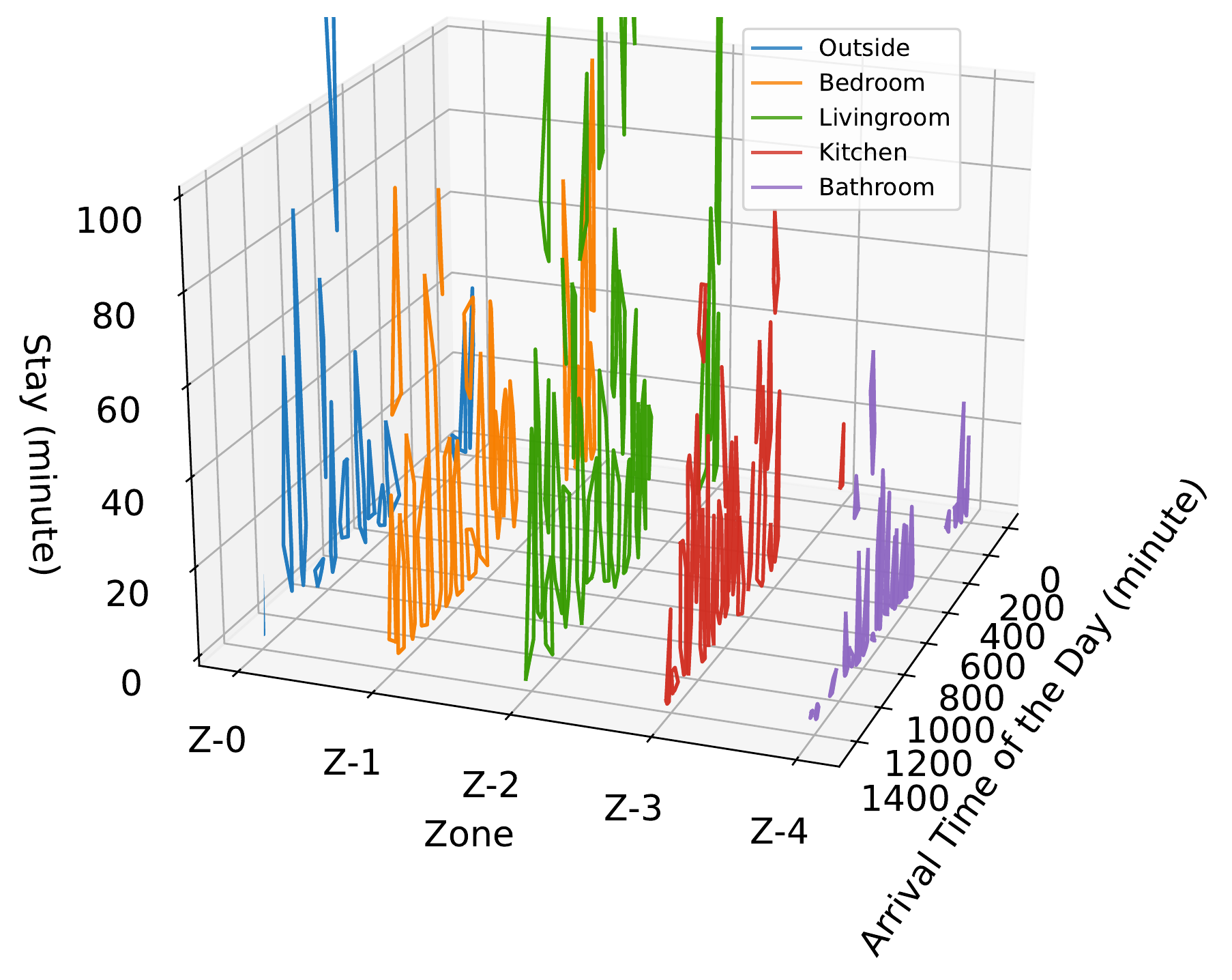}
        }     
    \end{center}
    \vspace{-10pt}
    \caption{Cluster visualizations of (a) DBSCAN-based ADM and (b) K-Means clustering-based for ARAS HAO1 dataset.}
    \label{fig:cluster_visualization}
    \vspace{-10pt}
\end{figure}

\noindent \textbf{Anomaly Detection Model (ADM)}
We consider ARAS datasets for evaluating our work, which captures every minute data of 27 different occupant activities from 4 zones of 2 homes (2 occupants each) over a month period~\cite{alemdar2013aras}. The dataset is used to train two different clustering-based ADMs- DBSCAN and K-Means clustering~\cite{ccelik2011anomaly, hartigan1979algorithm}. For training the ADMs, we consider four datasets, which we name HAO1 (i.e., a dataset containing information about one occupant of house A), HAO2, HBO1, and HBO2 datasets, and use the names throughout the write-up. The hyperparameter of the ADMs are optimized using Davies-Bouldin Index (DBI), Silhouette Coefficient (SC), and Calinski-Harabasz Index (CHI) since the ground truth of the clusters are not known~\cite{hubert1985comparing}. The higher values of SC and CHI and the lower value of DBI yield better performance. Figure~\ref{fig:cluster_visualization_} shows the performance of the ADMs based on different hyperparameters for HAO1. The optimal DBSCAN hyperparameter minPts (i.e., the minimum number of points for cluster forming) is found to be 30, where the optimal K-means clustering hyperparameter k (number of clusters) is 29. The other hyperparameter for DBSCAN (i.e., maximum distance in between within cluster samples) is considered to be 3 (i.e., the minimum number of points to create a convex hull). Since the datasets lack adequate samples, the ADMs' performance is not significant. However, the progressive learning capability for both ADMs (i.e., linearized with convex hull) shown in Figure~\ref{fig:anomaly_detection_progressive_performance} suggests that after learning a few more days/months of data, the ADM will fully learn the occupants' behavioral patterns. To evaluate the performance of the ADMs, we generated attack samples using the BIoTA framework~\cite{haque2021biota}. We use the F1-score (i.e., the harmonic mean of precision and recall) for the performance evaluation since the datasets are imbalanced datasets. The HAO1 dataset has 12.4\%, 12.1\%, 13.6\%, and 14.3\% attack data compared to the benign data for 10, 15, 20, and 25 days of training samples(out of 30 days), respectively. For the HAO2 dataset, the ratios are 3.6\%, 3.61\%, 3.73\%, and 3.71\%, respectively. Similarly, for HBO1 and HBO2 datasets, the ratios are around 7\%. The ADM clusters visualized in Figure~\ref{fig:cluster_visualization} show that the clusters from K-means clustering cover a larger area than DBSCAN clustering. The main reason is that the K-means clustering algorithm clusters every sample in training sets into benign samples (i.e., no benign or outlier samples). Here we mainly discuss the choice of ADMs and their hyperparameters. The performance of ADMs is evaluated, reasoned, and discussed more through the SHATTER framework in section~\ref{sec:evaluation}.

\subsection{Attack Model}
The attack model is used to generate parameterized attack procedures and functions that target a specific cyber-physical system (CPS), in our case, a smart home. In this section, we provide a summarized version of the attack model, which is detailed and formally analyzed in Section~\ref{sec:technical-details}.
%
%
\subsubsection{Attack Assumptions}
The proposed framework considers a set of assumptions.
\begin{compactenum}[(a)]
    \item \textbf{Assumption I:} Attacker has complete knowledge of the zone properties, smart home control algorithm, and ADM. Moreover, the parameters considered in the zone, appliance, and activity modeling are known to the attacker.
    \item \textbf{Assumption II:} We assume each zone of our considered smart home accommodates only single measurements for measuring the IAQ.
    \item \textbf{Assumption III:} We consider the attacker having access to sensor measurement (IAQ, occupancy, and appliances' status) can read and alter the measurement, while access to an appliance indicates the feasibility of activation of an unacitvated appliance.
    \item \textbf{Assumption IV:} All actuation devices cannot be altered or activated similar to smart appliances. For instance, cooling/heating fans and vents are out of the attack scope. 
    \item \textbf{Assumption V:} The controller and communication between the controller to the actuator is out of attack scope. Since the controllers are high computation devices they are hard and expensive parts to be compromised. Furthermore, the HVAC controllers and actuators are physically connected through a wired medium, which makes them sturdy against attackers' manipulation~\cite{belimo2023}.
\end{compactenum}
%

\subsubsection{Attack technique}

In our formal threat analysis framework, we are considering FDI or measurement manipulation attacks. Altering the sensor measurements, an attacker can lead the $\mathbb{E}$ to make an erroneous system state, thus making the $\mathbb{C}$ send an improper control signal, resulting in actuating the $\mathcal{A}$ differently than required. The measurement alternations are considered to be performed intelligently to evade the $\mathbb{E}$. The proposed framework finds out only those attack vectors (each contains the false values to be injected in different sensor measurements) that are attainable with the attacker's capability. It is to be noted that inaudible voice commands are also considered FDI in the attack model and are part of the attack vector. The $\mathbb{H}$ can be attacked with FDI attacks  in different ways. For instance, an attacker can leverage physical interaction features on IoT devices to conduct a stealthy attack against IoT systems. The attacks are primarily launched using an app or environment where an attacker seeks to exploit aphysical channels~\cite{ding2018safety}. The SHATTER-considered attacks can be broadly classified into two categories- sensor measurement acquisition and alteration as detailed followingly.

\noindent\textbf{{Measurement acquisition through physical sensing}}
The occupants' location (i.e., within $\mathcal{Z}$) can be sniffed by RF signals as depicted in~\cite{zhu2018tu}. The Access point, along with the $\mathcal{S}$ emits RF that constantly reflects on the occupant's body and, therefore, sends out the information about the occupant's location in the building. Here, the adversary uses commodity and low-cost sniffers to conduct a covert reconnaissance attack that can continually monitor and pinpoint human activity within a particular location in a home or an office without having any physical or remote access to the WiFi devices.

\noindent\textbf{{Measurement acquisition through eavesdropping communication packets}}
In an IoT network, nodes generally send, forward, or collect packets along with evaluating the routing consistency of each path~\cite{liu2021detection}. The attacker's access to the router can act as a man-in-the-middle and sniff the communication packets through packet capture and analysis tools.

\noindent\textbf{{Measurement alteration through packet crafting}}
Through man-in-the-middle attack, the attacker can not only eavesdrop on packets but also alter/craft packet information utilizing ARP poisoning and IP/MAC addressing spoofing attack. The goal is to alter the measurement information sent by specific sensors. Through this approach, the attacker can modify all the sensor measurements and appliances' statuses. Such alterations are hardly detected by the control system. 

\noindent\textbf{{Activation of Appliances through inaudible voice commands}} Smart IoT devices require charging whenever the battery becomes low in other to function appropriately. A  recent attack has shown the feasibility of sending inaudible voice commands from smartphones through malicious charging plugs~\cite{ding2021iotsafe}. To perform this attack, the occupant must charge their device. Moreover, inaudible voice commands can be transmitted through other approaches as identified by the existing works -- Backdoor~\cite{roy2017backdoor}, DolphinAttack~\cite{zhang2017dolphinattack}, LipRead~\cite{roy2018inaudible}, SurfingAttack~\cite{yan2020surfingattack}, etc. These attacks allow an adversary to send inaudible voice commands to the voice assistants and stealthily activate the appliances.
\subsubsection{Attack Goal}

The SHATTER framework's primary goal is to increase the energy consumption of the home through stealthy FDI attacks. The attack demands alteration of the necessary sensor measurements to maximize the overall energy consumption in $\mathcal{H}$ by forcing $\mathcal{C}$ to flow more air (i.e., both fresh and return air) through the supply air duct in different $\mathcal{Z}$. To launch stealthy attacks, the attacker needs to bypass:

\noindent\textbf{ADM:} Inconsistencies in the occupancy or IAQ measurement from the learned occupant's behavioral pattern will be recognized as an anomalous event.

\noindent\textbf{Occupants:} Turning on the washer in the kitchen, while an occupant is cooking, will lose the attack's stealthiness.

\subsubsection{Attacker's Attributes}
For modeling the attack, we consider variable accessibility and resource constraints for the attacker. This work considers a knowledgeable attacker aware of the surrounding weather pattern, smart home zone attributes, occupancy information, underlying control, and defense mechanisms (i.e., ADM) of the smart home control system. It is unreasonable to assume that attacker has access accessibility to all of the resources (i.e., sensor devices) to initiate a stealthy FDI attack. The attack model specifies access to sensor measurements (i.e., IAQ/occupancy/appliances' status measurement), and access to appliances (i.e., appliances that can be triggered by inaudible voice commands).
%
The major differences between BIoTA and SHATTER framework are illustrated in Table~\ref{tab:biota_vs_shatter}.

\begin{table}[t]
\caption{Primary differences between BIoTA and SHATTER framework.}
\label{tab:biota_vs_shatter}
\vspace{-6pt}
\begin{tabular}{|p{2.3cm}|p{2.7cm}|p{2.7cm}|}
\hline
\textbf{Criteria}                                                               & \textbf{Bthe IoTA}                                                                                                      & \textbf{SHATTER}                                                                                                                      \\ \hline
\textbf{Application Domain}                                                     & Smart building/ homes                                                                                               & Smart homes                                                                                                                           \\ \hline
\textbf{Number of Occupants}                                                    & $\sim$(10-1000)                                                                                                     & $\sim$(1-10)                                                                                                                          \\ \hline
\textbf{\begin{tabular}[c]{@{}l@{}}Anomaly Detection\\ Model\end{tabular}}      & Rule-based                                                                                                          & ML-based                                                                                                                \\ \hline
\textbf{\begin{tabular}[c]{@{}l@{}}Occupant's Activity\\ Tracking\end{tabular}} & Not considered                                                                                                      & Considered                                                                                                                         \\ \hline
\textbf{Appliance Modeling}                                                     & Fixed load at every control cycle                               & Dynamic load modeling
\\ \hline
\textbf{Attack Constraints}                                                     & Stealthy bypass control system                              & Deceive both control system and occupants                             \\ \hline

\textbf{Attack Technique}                                                            & Greedy FDI attack & Dynamic FDI attack                              \\ \hline
\end{tabular}
\vspace{-12pt}
\end{table}

%
\section{Technical Details of the SHATTER Framework}
\label{sec:technical-details}
In this section, we provide a detailed overview of the SHATTER framework. SHATTER formally models the smart home control systems, ADM, and the attack model using the Z3 tool that leverages SMT~\cite{de2008z3}-based solver and optimizer to identify stealthy attack vectors that can optimally increase the energy consumption of the home. Table~\ref{tab:modeling-notations} demonstrates the modeling notations.

\vspace{-7pt}
\subsection{Formal Modeling of the smart home control system}
For the HVAC control system, we mainly consider temperature, occupancy, and $CO_2$ concentration as measurement values. Because building occupants are the primary source of continuous heat and $CO_2$ generation, accurately measuring the number of people in real-time utilizing different building sensor systems is critical for computing energy efficiency and occupant comfort.

\begin{table}[!t]
\small
\centering
\caption{Modeling Notations} 
\label{tab:modeling-notations}
\vspace{-6pt}
\scriptsize
\begin{tabular}{|p{1.3cm}|p{0.5cm}|p{4.5cm}|p{0.7cm}|}
\hline
\textbf{Type of Notation}               & \multicolumn{1}{c|}{{\begin{tabular}[c]{@{}l@{}}\textbf{Nota-} \\ \textbf{tion}\end{tabular}} }                  & \multicolumn{1}{c|}{\textbf{Description}}                                                                                                                                              & \multicolumn{1}{c|}{\begin{tabular}[c]{@{}l@{}}\textbf{Data} \\ \textbf{Type}\end{tabular}} \\ \hline
\multirow{8}{*}{General}                & $\mathcal{Z}$                                           & Set of all zones                                                                                                                                                                       & Set                                     \\ \cline{2-4} 
                                        & $\mathcal{O}$                                           & Set of all occupants in a multi-occupant setup                                                                                                                                      & Set                                     \\ \cline{2-4} 
                                        & $\mathcal{D}$                                           & Set of all appliances                                                                                                                                                                  & Set                                     \\ \cline{2-4} 
                                        & $\mathcal{T}$                                           & Set of all timeslots in a day at each sampling time                                                                                                                                    & Set                                     \\ \cline{2-4} 
                                        & $\mathcal{A}$                                           & Set of all activities                                                                                                                                                                  & Set                                     \\ \cline{2-4} 
                                        & $\mathcal{D}_{z,d}$                                     & d-th appliance at zone, $z$                                                                                                                                                            & Integer                                 \\ \cline{2-4} 
                                        & $\mathcal{A}_\mathit{t,o,z}$                            & Activity conducted by $o$-th occupant at $t$-th timeslot in $z$-th zone                                                                                                                              & Integer                                 \\ \cline{2-4} 
                                        & $\Delta t$                                              & Sampling time of the controller                                                                                                                                                        & Integer                                 \\ \hline
\multirow{5}{*}{\begin{tabular}[c]{@{}l@{}}Sensor \\ Measurements\end{tabular}}   & $\mathcal{S}$                                           & Set of all sensor measurements                                                                                                                                                         & Set                                     \\ \cline{2-4} 
                                    & $\mathcal{S}^{OE}_{t,z}$                               & Occupancy estimation measurement (occupants count) at $t$-th timeslot in $z$-th zone                                                                                                                       & Integer                                 \\ \cline{2-4}
                                        & $\mathcal{S}^{OT}_{t,o,z}$                               & Tracking presence of $o$-th occupant at $t$-th timeslot in $z$-th zone                                                                                               & Boolean                                 \\ \cline{2-4} 
                                        & $\mathcal{S}^{C}_{t,z}$                                 & $CO_2$ sensor measurement at $t$-th timeslot in $z$-th zone                                                                                                                           & Real                                    \\ \cline{2-4} 
                                        & $\mathcal{S}^{T}_{t,z}$                                 & Temperature ($^{\circ}$ F) sensor measurement at $t$-th timeslot in $z$-th zone                                                                                                            & Real                                    \\ \cline{2-4} 
                                        & $\mathcal{S}^{D}_{t,z,d}$                               & d-th appliance's status (on/off) at $t$-th timeslot in $z$-th zone                                                                                                                          & Boolean                                 \\ \hline
\multirow{2}{*}{\begin{tabular}[c]{@{}l@{}}Actuation \\ Measurements\end{tabular}} & $\mathcal{Q}$                                           & Set of all airflow (cfm) measurements                                                                                                                                                  & Set                                     \\ \cline{2-4} 
                                        & $\mathcal{Q}_{t, z}$                                 & Airflow based at $t$-th timeslot in $z$-th zone  & Real                                    \\ \hline
\multirow{4}{*}{\begin{tabular}[c]{@{}l@{}}Variable \\ Parameters\end{tabular}}    & $\mathbb{P}^{OT}_{t}$                                   & Outdoor temperature at timeslot, $t$                                                                                                                                                   & Real                                    \\ \cline{2-4} 
                                        & $\mathbb{P}^{OC}_{t}$                                   & Outdoor $CO_2$ concentration at $t$-th timeslot                                                                                                                                         & Real                                    \\ \cline{2-4} 
                                        & $\mathbb{P}^{CS}_\mathit{t, z}$                         & $CO_2$ setpoint at $t$-th timeslot in $z$-th zone                                                                                                                                           & Real                                    \\ \cline{2-4}
                                        & $\mathbb{P}^{TSP}_\mathit{t, z}$                         & Temperature of supply air at $t$-th timeslot in $z$-th zone                                                                                                                                           & Real                                    \\ \cline{2-4}
                                        & $\mathbb{P}^{TS}_\mathit{t, z}$                         & Temperature setpoint at $t$-th timeslot in $z$-th zone                                                                                                                                    & Real                                    \\
                                        \cline{2-4}
                                        & $\mathbb{P}^{TM}_\mathit{t, z}$                         & Temperature of mixed air at $t$-th timeslot in $z$-th zone                                                                                                                                    & Real                                    \\
                                                                                \cline{2-4}
                                    
                                                                                \cline{2-4}
                                        & $\mathbb{P}^{TEC}_{t}$                         & Total energy consumption (kWh) at $t$-th timeslot  & Real                                    \\
                                        \hline
\multirow{4}{*}{\begin{tabular}[c]{@{}l@{}}Fixed \\ Parameters\end{tabular}}         & $\mathbb{P}^{CE}_{o,z,a}$                               & $CO_2$ emission per person per minute for occupant $o$ at $z$-th zone performing activity, $a$                                                                                         & Real                                    \\ \cline{2-4} 
                                        & $\mathbb{P}^{HR}_{o,z,a}$                               & Heat radiation per person per minute for $o$-th occupant at $z$-th zone performing $a$-th activity                                                              & Real                                    \\ \cline{2-4} 
                                        & $\mathbb{P}^{V}_{z}$                                    & Volume ($ft^3$) of zone, $z$                                                                                                                                                           & Real                                    \\ \cline{2-4} 
                                        & $\mathbb{P}^{PC}_{d}$                                   & Power consumption (Watt) of the d-th appliance if it is turned on                                                                                                                        & Real                                    \\
                                        \cline{2-4} 
                                        & $\mathbb{P}^{HRF}_{d}$                                   & Heat radiation factor of d-th appliance that nees to be multiplied by power consumption (Watt) to obtain the heat radiation (kWh) from appliance                                                                                                                     & Real                                    \\
                                        \cline{2-4} 
                                        & $\mathbb{P}^{COP}_t$                                   & Off-peak hour energy cost (\$/kWh)                                                                                                            & Real                                    \\
                                        \cline{2-4} 
                                        & $\mathbb{P}^{CP}_t$                                   & Peak hour energy cost (\$/kWh)                                                                                                            & Real                                    \\
                                        \cline{2-4} 
                                        & $\mathbb{P}^{BS}$                                   & Battery total storage (kWh) that is charged at off peak hours and used at peak hour to reduce peak hour energy cost                                                                                                               & Real                                    \\
                                        \hline
\multirow{4}{*}{\begin{tabular}[c]{@{}l@{}}Attack \\ Vector\end{tabular}}           & $\delta_{t,z}^{C}$   & False measurement to be added in $CO_2$ sensor measurements in zone, $z$ at timeslot $t$                                                                                               & Real                                    \\ \cline{2-4} 
                                        & $\delta_{t,z}^{T}$   & False measurement to be added in temperature sensor measurementat $t$-th timeslot in $z$-th zone                                                                                          & Real                                    \\ \cline{2-4} 
                                        & $\delta_{t,o, z}^{O}$ & False measurement to be multiplied with occupancy sensor measurements for $o$-th occupant at $t$-th timeslot in $z$-th zone                       & Boolean                                 \\ \cline{2-4} 
                                        & $\delta_{t,z,d}^{D}$ & False measurement to be multiplied with $d$-th appliance sensor measurements at $t$-th timeslot in $z$-th zone               & Boolean                                 \\
                                        \cline{2-4} 
                                        & $\mathcal{I}$ & Attack optimization window               & Integer                                 \\\hline
\end{tabular}
\vspace{-25pt}
\end{table}

\textbf{Ventilation control Constraints:}
The HVAC control system adds optimal fresh outside air to the supply air for maintaining the $CO_2$ concentration in occupants' comfort range. The ventilation requirement depends on $CO_2$ emitted by the occupants, which varies based on the metabolic rate (depending on the occupants' age and levels of conducted physical activities).
\begin{equation}
    \label{eq:ventilation_control}
    \begin{split}
    \forall_{t \in \mathcal{T}} \forall_{z \in \mathcal{Z}} ~~\frac{\mathcal{S}^{OE}_{t,z} \times \mathbb{P}^{\mathit{CE}}_{o,z,a = \mathcal{A}_{t,o,z}} \times \Delta t}{\mathbb{P}^{V}_{z}} = \mathbb{P}^{\mathit{CS}}_{\mathit{t,z}}~-\\ \left(1-\frac{\mathcal{Q}_{t, z}}{\mathbb{P}^{\mathit{V}}_{z}}\right) \times
    \mathcal{S}^\mathit{C}_{t, z} 
    - \frac{\mathcal{Q}_{t,z} \times \Delta t}{\mathbb{P}^{V}_{z}} \times \mathbb{P}^{\mathit{OC}}_{t}
    \end{split}
\end{equation}
%

\textbf{Temperature Control Constraints:}
The HVAC control system optimizes the usage of zone return air to the supply air for quickly meeting the zones' setpoint temperature and minimizing the home's energy consumption. The cooling demand is dependent on the appliances' heat radiation and the occupants' metabolic rate. In this work, we consider the load demand based on the appliances' status, unlike the control rules of BIoTA (constant load demand). The factor 0.3167 in Equation \ref{eq:temperature_control} is used since it does not vary significantly with the parameters change. We multiply a factor ($\mathbb{P}^{HRF}_{d}$) with total energy consumption for all devices to calculate sensible heat gain (e.g., LED lights radiate 12\% heat~\cite{ledlightinginfo2020}).
%
\begin{equation}
    \label{eq:temperature_control}
    \begin{split}
    \forall_{t \in \mathcal{T}} \forall_{z \in \mathcal{Z}} \forall_{d \in \mathcal{D}} ~~ {\mathcal{Q}_{t, z} \times (\mathbb{P}^{\mathit{TS}}_{z, t} -\mathbb{P}^{\mathit{TSP}}_{z, t})} \times 0.3167 = \\ \mathcal{S}^{OE}_\mathit{t,z} \times 
     \mathbb{P}^{\mathit{HR}}_{o, z, a = \mathcal{A}_{t,o,z} } + \mathcal{S}^{D}_{t,z,d} \times \mathbb{P}^{\mathit{PC}}_{d} \times \mathbb{P}^{HRF}_{d}
    \end{split}
\end{equation}
%
Equations~\ref{eq:ventilation_control} and~\ref{eq:temperature_control} constraint the airflow requirement in the zones and account for optimal estimation of airflow to satisfy both occupant's comfort and energy savings needs.

\textbf{HVAC Control Cost Calculation:}
The HVAC cost calculation mainly depends on the air quality in the mixed air chamber. The air handling unit (AHU) optimally mixes the fresh and return air to meet the zone's IAQ demand. The air mixing chamber of the HVAC controller mixes both outside fresh air and recirculating return air optimally to meet energy efficiency and occupants' comfort. 
\begin{equation}    \label{eq:instantaneous_total_energy_consumption}
    \begin{split}
    \forall_{t \in \mathcal{T}} \mathbb{P}^{\mathit{TEC}}_{\mathit{t}} = \sum_{z \in \mathcal{Z}} ~~ \mathcal{Q}_{t, z} \times (\mathbb{P}^{\mathit{TM}}_{z, t} -\mathbb{P}^{\mathit{TSP}}_{z, t}) \times 0.3167 \\ \times \frac{\Delta t}{60000} + \sum_{z \in \mathcal{Z}, d \in \mathcal{D}}\mathcal{S}^{D}_{t,z,d} \times \mathbb{P}^{\mathit{PC}}_{d}
    \end{split}
\end{equation}

\begin{equation}
    \label{eq:energy_cost}
    \begin{split}
    \mathcal{G}^{\mathcal{S}} = \sum_{\mathit{t_1} \in \mathcal{T}^{\mathit{OP}} \lor (\mathit{t_1} \in \mathcal{T}^{P} \land \sum_{t = \mathcal{T}_0^{P}}^{\mathit{t_1}} \mathbb{P}^{\mathit{TEC}}_{t} \leq \mathbb{P}^{\mathit{BS}})} ~~ \mathbb{P}^{\mathit{TEC}}_{\mathit{t_1}} \times \mathbb{P}^{\mathit{COP}}_{\mathit{t_1}}  \\ + \sum_{ (\mathit{t_2} \in \mathcal{T}^{P} \land \sum_{t = \mathcal{T}_0^{P}}^{\mathit{t_2}} \mathbb{P}^{\mathit{TEC}}_{\mathit{t_2}} > \mathbb{P}^{\mathit{BS}})} \mathbb{P}^{\mathit{TEC}}_{\mathit{t_2}} \times \mathbb{P}^{\mathit{CP}}_{\mathit{t_2}} 
    \end{split}
\end{equation}
%
%
The instantaneous power consumption considers both HVAC and appliance-induced consumption as shown in Equation~\ref{eq:instantaneous_total_energy_consumption}. We assume that the home has battery storage that is charged at the off-peak hour and discharged at the peak hour to meet its energy demand and thus reduce the household energy cost. The energy pricing is taken from PG\&E electricity rate plans~\cite{pse2023}. For brevity, we consider that the battery storage is always charged the full during off-peak hours. Hence, during off-peak hours and a portion of peak hours (i.e., until the battery is fully discharged), the residential loads operate at off-peak hour costing as shown in Equation~\ref{eq:energy_cost}.
\subsection{Formal Modeling of the Anomaly Detection Model (ADM)}
The SHATTER framework also extracts formal constraints from the ADM to generate stealthy attack vectors. We consider a clustering-based anomaly detection approach in this work, which continuously checks the duration of stay for an occupant in a particular zone based on the arrival time of that occupant. The considered ADM uses a clustering technique to attain the valid pairs of (arrival time and duration of staying). The hypothesis of choosing this approach is that occupants converge to a set of actions (i.e., moving from one zone to another, doing household chores) after habit formation. In the following write-up, we formally model the ADM after providing an intuitive explanation of the ADM using a toy example. 
We consider that ADM always checks the duration of staying, $t_2$, at a particular zone, while the occupant has entered the zone at the time, $t_1$, with a pre-trained model. The pre-trained model comes up with several clusters. If the point ($t_1$, $t_2$) is not within any of the clusters, the controller raises the alarm.

\begin{figure}[!t]
    \centering
    \includegraphics[width=0.75\columnwidth]{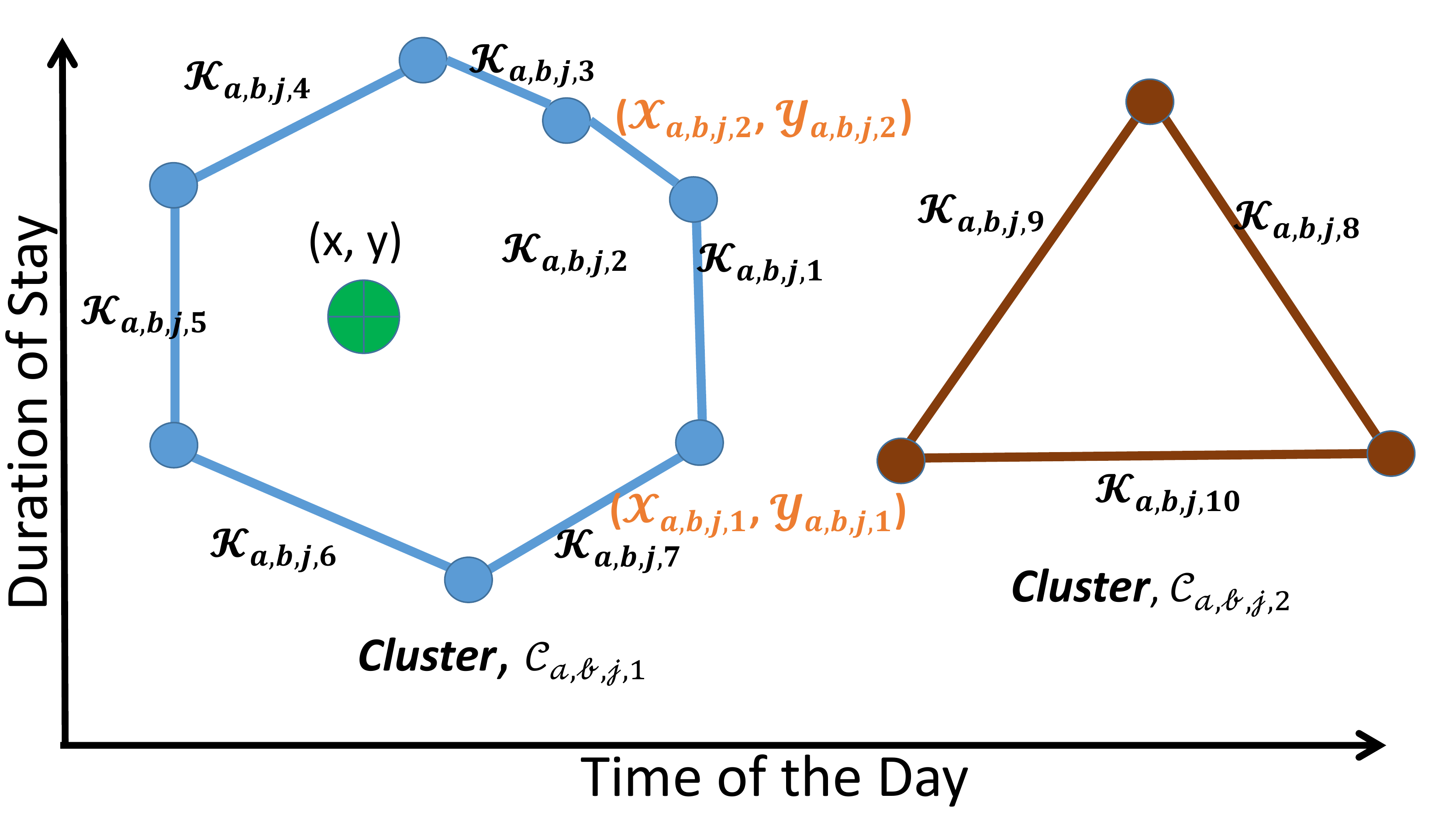}
    \vspace{-12pt}
    \caption{Sample convex hull representation of the ADM cluster with formal notations.}
    \vspace{-12pt}
    \label{fig:toy_example}
\end{figure}

Figure~\ref{fig:toy_example} shows two clusters ($\mathcal{C}_{\mathit{o,z,1}}$ and ($\mathcal{C}_{\mathit{o,z,2}}$) in a 2D data plane where $\mathcal{C}_{\mathit{o,z,1}}$ consists of seven line segments ($\mathcal{K}_{\mathit{o,z,1}}$, $\mathcal{K}_{\mathit{o,z,2}}$, \ldots, $\mathcal{K}_{\mathit{o,z,7}}$) and $\mathcal{C}_{\mathit{o,z,2}}$ consists of three line segments ($\mathcal{K}_{\mathit{o,z,8}}$, $\mathcal{K}_{\mathit{o,z,9}}$, and $\mathcal{K}_{\mathit{o,z,10}}$). We denote the end points of any line segment ($\mathcal{K}_{\mathit{o,z,i}}$) are ($\mathcal{X}_{\mathit{o,z,i}}, \mathcal{Y}_{\mathit{o,z,i}}$) and 
($\mathcal{X}_{\mathit{o,z,i}}, \mathcal{Y}_{\mathit{o,z,i}}$), where 
$\mathcal{Y}_{\mathit{o,z,i}} \geq \mathcal{Y}_{\mathit{o,z,i}}$ .
%
\begin{compactenum}[(a)]
\item $\mathit{leftOfLineSegment}(t_1, t_2, \mathcal{K}_{\mathit{o,z,i}})$: This function checks if the point $(t_1,t_2)$ is on the left side of the line segment, $\mathcal{K}_{\mathit{o,z,i}} $.
\item $\mathit{withinCluster}(t_1, t_2, \mathcal{C}_{\mathit{o, z, k}})$: This function returns $True$ if the data point $(t_1,t_2)$ is within the cluster, $\mathcal{C}_{\mathit{o, z, k}}$. The point is considered to be within a cluster if it is $\mathit{leftOfLineSegment}$ of all the clusters.
\end{compactenum}

A new set of formal modeling notations- $\mathcal{E}_{t_1,o,z}^{A}$, $\mathcal{E}_{t_1,o,z}^{E}$, and $\mathcal{E}_{t_1,o,z}^{S}$ derived from the occupancy sensor measurements for modeling the ADM. As the name suggests, $\mathcal{E}^{\mathit{A}}$ and $\mathcal{E}^{E}$ respectively denote the arrival and exit events of all zones and for all occupants.
\begin{equation}\label{eq:arrival_event}
    \begin{split}
     \forall_{t_1 \in \mathcal{T}} ~~\mathcal{E}_{t_1,o,z}^{A} \rightarrow \mathcal{S}^{OT}_{t_1,O,z} \land \neg \mathcal{S}^{OT}_{t_1 - 1,O,z}
\end{split}
\end{equation}
%
\begin{equation}\label{eq:exit_event}
    \begin{split}
     \forall_{t_2 \in \mathcal{T}} ~~\mathcal{E}_{t_2,o,z}^{E} \rightarrow \ \mathcal{S}^{OT}_{t_2,o,z} \land  \neg \mathcal{S}^{OT}_{t_2 + 1,o,z}
\end{split}
\end{equation}

The stay duration for the occupants at a specific zone can be modeled using the arrival and exit events.
\begin{equation}\label{eq:event_stay}
    \begin{split}
     \forall_{t_1 \in \mathcal{T}} ~~\mathcal{E}_{t_1,o,z}^{\mathit{S}} = (t_2 - t_1) \rightarrow \mathcal{E}_{t_1,o,z}^{\mathit{A}} \land \mathcal{E}_{t_2> t_1,o,z}^{\mathit{E}}  \\ \land \forall_{t_1 < t_3 < t_2 \in \mathcal{T}} \mathcal{S}^{OT}_{t_3,o,z}
\end{split}
\end{equation}

The occupancy sensor measurements are considered to be benign if all the entering and leaving events of the occupants are consistent with the DBSCAN clusters.
\begin{equation}\label{eq:consistent}
\begin{split}
\mathit{consistent}(\mathcal{S}^{\mathit{OT}}) \rightarrow \forall_{o \in \mathcal{O}} \forall_{z \in \mathcal{Z}} \forall_{t_1 \land \mathcal{E}_{t_1,o,z}^{\mathit{E}}} \\ ~~ \mathit{withinCluster}(t_1, t_2 = \mathcal{E}_{t_1,o,z}^{\mathit{S}}, \mathcal{C}_{z,o})
\end{split}
\end{equation}

Here,
\begin{equation}
    \label{eq:within_cluster}
    \begin{split}
     & \mathit{withinCluster}(t_1, t_2, \mathcal{C}_{z,o}) \rightarrow \\
     & ~~ \exists_{c \in \mathcal{C}_{z,o}}
     \forall_{k \in {\mathcal{K}_{z,o} \land In(k, c)}} ~~ \mathit{leftOfLineSegment}(t_1,t_2, k)
\end{split}
\end{equation}

\begin{equation}\label{eq:left_of_line_segment}
\begin{split}
&\mathit{leftOfLineSegment}(t_1, t_2 = \mathcal{E}_{t_1,o,z}^{\mathit{S}}, \mathcal{K}_{z,o, i}) \rightarrow \\
&
(t_1(\mathcal{Y}_{z,o,i} - \mathcal{Y}_{z,o,i}) - t_2(\mathcal{X}_{z,o,i} - \mathcal{X}_{z,o,i}) -  (\mathcal{X}_{z,o,i} \mathcal{Y}_{z,o,i}  \\ & - \mathcal{X}_{z,o,i}  \mathcal{Y}_{z,o,i})) < 0
\end{split}
\end{equation}

The equations~\ref{eq:within_cluster} and~\ref{eq:left_of_line_segment} say that the ($t_1$, $t_2$) pairs will be considered to be within the valid clusters if at least one cluster encloses the point. The condition of being in a cluster (i.e., convex hull) is that the point is the left-hand side of all that cluster line segments.

\subsection{Formal Modeling of Attacks}
\label{subsec:formal_attack_model}
The main goal of the attack is to maximize the energy cost by adding false measurements.  The following three equations demonstrate the FDI attack in IAQ, occupancy, and appliance measurements, respectively.
\begin{compactenum}[(1)]
    \item $\forall_{t \in \mathcal{T}^{A}}\forall_{z \in \mathcal{Z}^{A}}\forall_{p \in [\mathit{C, T}]} \mathcal{\bar{S}}_{t,z}^{\mathit{p}} = \mathcal{S}_{t,z}^{p} + \delta_{\mathit{t,z}}^{p}$
    \item $\forall_{t \in \mathcal{T}^{A}} \forall_{o \in \mathcal{O}^{A}} \forall_{z \in \mathcal{Z}^{A}} \mathcal{\bar{S}}_{t,o,z}^{\mathit{OT}} =  \mathcal{S}_{t,o,z}^{\mathit{OT}} \times \delta_{t,o,z}^{\mathit{OT}}$
    \item $\forall_{t \in \mathcal{T}^{A}} \forall_{z \in \mathcal{Z}^{A}} \forall_{d \in \mathcal{D}^{A}} \mathcal{\bar{S}}_{t,z,d}^{\mathit{D}} =  \mathcal{S}_{t,z,d}^{\mathit{D}} \times \delta_{t,z,d}^{\mathit{D}}$
\end{compactenum}
Here, $\delta$ is the attack vector that denotes the required injection to accomplish the attack goal.

\noindent\textbf{Attack Goal:}
\begin{equation}
\label{eqn:attack_goal}
maximize ~\mathcal{G}^{\mathcal{\bar{S}}}
\end{equation}
%
\noindent\textbf{Attack Constraints:}
\begin{equation}
\label{eqn:consistent_delta}
\begin{split}
\mathit{consistent(\mathcal{S}^{\mathit{OT}} + \delta^{\mathit{OT}})}
\end{split}
\end{equation}
\begin{equation}
\label{eqn:ap}
\begin{split}
\sum_{t \in \mathcal{T}^{A}, o \in \mathcal{O}^{A}, z \in \mathcal{Z}^{A}} \mathcal{\bar{S}}_{t,o,z}^{\mathit{OT}} = \sum_{t \in \mathcal{T}^{A}, o \in \mathcal{O}^{A}, z \in \mathcal{Z}^{A}} \mathcal{S}_{t,o,z}^{\mathit{Occ}} 
\end{split}
\end{equation}

\begin{equation}
    \label{eq:ventilation_verification}
    \begin{split}
    & \forall_{t \in \mathcal{T}^{A}} \forall_{o \in \mathcal{O}^{A}} \forall_{z \in \mathcal{Z}^{A}} ~ \frac{\mathcal{S}^{OE}_{t - 1,z} \times \mathbb{P}^{\mathit{CE}}_{o,z,a = \mathcal{A}_{t - 1,o,z}} \times \Delta t}{\mathbb{P}^{V}_{z}}  =  \mathcal{S}^{C}_{t, z} \\ & - \left(1- \frac{\mathcal{Q}_{t - 1, z}}{\mathbb{P}^{\mathit{V}}_{z}}\right)  \times
    \mathcal{S}^{C}_{t - 1, z}
    - \frac{\mathcal{Q}_{t - 1,z} \times \Delta t}{\mathbb{P}^{\mathit{V}}_{z}} \times \mathbb{P}^{\mathit{OC}}_{t - 1}
    \end{split}
\end{equation}

\begin{equation}
    \label{eq:temperature_verification}
    \begin{split}
    & \forall_{t \in \mathcal{T}^{A}} \forall_{z \in \mathcal{Z}^{A}} \forall_{d \in \mathcal{D}^{A}} ~~ {\mathcal{Q}_{t- 1, z}  \times (\mathcal{S}^{T}_{t, z} - \mathcal{S}^{T}_{t - 1, z})} \\ & \times 0.3167 = \mathcal{S}^{OE}_\mathit{t,z} \times \mathbb{P}^{\mathit{HR}}_{o, z, a = \mathcal{A}_{t,o,z} } +\mathcal{S}^{D}_{t,z,d} \times \mathbb{P}^{\mathit{PC}}_{d} \times \mathbb{P}^{HRF}_{d}
    \end{split}
\end{equation}

\begin{equation}
    \label{eq:appliance_verification}
    \begin{split}
    \forall_{t \in \mathcal{T}^{A}} \forall_{z \in \mathcal{Z}^{A}} \forall_{d \in \mathcal{D}^{A}} ~~\mathcal{\bar{S}}_{t,z,d}^{D} = \neg\mathcal{S}_{t,z,d}^{D} \rightarrow \forall_{o \in \mathcal{O}} \mathit{stealthy}(\mathit{d, o})
    \end{split}
\end{equation}

These are attack constraints, where Equation~\ref{eqn:consistent_delta} demands that altered occupancy measurement should follow the clusters. Equations~\ref{eq:ventilation_verification} and~\ref{eq:temperature_verification} are adopted from the BIoTA framework, which requires prediction made in the previous timeslot about sensor measurements and actuation should be consistent with the current timeslot. The constraint in Equation~\ref{eq:appliance_verification} says that inaudible voice command-based appliance activation is possible if the device is present in an unoccupied zone.  

\noindent\textbf{Attacker's Property:}
An attacker may change a sensor measurement if he/she has the accessibility to that particular measurement. The attacker cannot inject false measurements into the inaccessible sensor measurements. The accessibility to zone, time-slots, devices, and occupants (RFID measurement) are modeled using $\mathcal{Z}^{A}$, $\mathcal{T}^{A}$, $\mathcal{D}^{A}$, and $\mathcal{O}^{A}$ respectively.

\noindent\textbf{Attack Technique:}
The principal task of the proposed attack is to misinform the controller with tailored occupants' location and activity information. Hence, the attack can be considered as a scheduling problem, where the attacker will compute an optimal schedule of occupants (along with the activities) throughout different zones at different time instances that evade both the ADM and occupants. Eventually, the optimization objective defined in Equation~\ref{eqn:attack_goal} is an NP-hard problem (i.e., complexity $\mathcal{O}(|\mathcal{Z}|^{|\mathcal{T}|})$). Hence, it is not feasible to get the optimal attack vectors in a viable time. SHATTER aims at identifying sub-optimal solutions by optimizing the scheduling problem in a shorter time window ($\mathcal{I}$) and merging the results. We will consider the schedule as an attack schedule throughout the write-up.

\begin{compactenum}[(a)]
    \item \textbf{Attack Schedule Generation:}
    In this process, the attacker pre-computes the attack schedule based on his knowledge of the control system and ADM. The goal of creating the attack schedule is to maximize the energy cost in the time horizon ($\mathcal{I}$), in which the optimization is feasible. 
    
    \noindent\textbf{Attack Schedule Goal:}
    \begin{equation}
    \label{eqn:attack_goal}
    \forall_{t \in [1, |\mathcal{T}|, \mathcal{I}]} ~~
    maximize \sum_{t}^{t + \mathcal{I}}\mathcal{G}^{\mathcal{\bar{S}}}_t
    \end{equation}
    However, the following attack constraints must be maintained to create the attack schedule.  
    \begin{equation}
    \label{eq:attack_schedule_zone_constraint}
        \begin{split}
        \forall_{t \in \mathcal {T}^{A}} \forall_{o \in \mathcal{O}^{A}} \exists!_{z \in \mathcal{Z}^{A}} \ ~~\mathcal{\bar{S}}_{t,o,z}^{\mathit{OT}}
        \end{split}
    \end{equation}
    %
    \begin{equation}
        \label{eq:attack_schedule_max_stay_constraint}
        \begin{split}
        \forall_{t \in \mathcal{T}^{A}} \forall_{o \in \mathcal{O}^{A}} \forall_{z \in \mathcal{Z}^{A}} ~~ \neg \mathcal{\bar{S}}_{t,o,z}^{\mathit{OT}} \rightarrow \mathcal{\bar{E}}_{\mathit{t-m,o,z}}^{A} \\ \land m = maxStay(t, o, z)
        \end{split}
    \end{equation}
    
    \begin{equation}
        \label{eq:attack_schedule_in_range_constraint}
        \begin{split}
        \forall_{t \in \mathcal{T}^{A}} \forall_{o \in \mathcal{O}^{A}} \forall_{z \in \mathcal{Z}^{A}} ~~  \mathcal{\bar{E}}_{\mathit{t,o,z}}^{E} \rightarrow \\ \exists_{x \in \mathcal{T}} ~~
        \mathit{inRangeStay} (t, o, z, \mathcal{\bar{E}}_{\mathit{t - x,o,z}}^{S})
        \end{split}
    \end{equation}

    Here, two functions are introduced. The maxStay(.) function outcomes the maximum valid stay duration (without alarming ADM) at a zone for an occupant given his/her arrival time. On the other hand, the inRangeStay(.) function checks whether staying at a zone for an occupant given his/her arrival time and stay duration is stealthy or not. The attack schedule can be derived from the attacked occupancy sensor measurements However, the attacker needs to make sure that the occupants are scheduled to a zone in every attack timeslot as shown in Equation~\ref{eq:attack_schedule_zone_constraint}. The equation~\ref{eq:attack_schedule_max_stay_constraint} requires that the attacker must schedule an occupant to a different zone if keeping the occupant more will alarm the ADM. Other than that the occupant can only be scheduled to a different zone if the stay duration in the current zone based on the arrival time is within an ADM cluster as shown in Equation~\ref{eq:attack_schedule_in_range_constraint}. Otherwise, scheduling the occupant in a different zone will alarm the ADM.
        \item \textbf{Real-timeAttack:} The pre-computed attack schedule can evade the ADM. However, to evade the occupants, the real-time measurement manipulation and appliance triggering decision need to be taken in real-time since the real-time occupant behavior will be different than the attack schedule. In real-time there will be two tasks - 1) using the attack schedule to measure manipulation and 2) appliance triggering attack. The former task requires misinforming the controller's IAQ and occupancy information according to the attack schedule. However, the attack can be carried out at a time-instances if the attacker has access to both the actual occupant zone and the zone from the attack schedule. The core idea behind the later task is that the appliances will be triggered based on the activity reported by the attack schedule, if and only if the occupant staying in the current zone has not exceeded the ADM reported minimum amount of time based on his/her arrival time. The algorithm of the appliance triggering process is shown in Algorithm~\ref{alg:appl_trig}, which sets a variable $trig$ to be $True$ when adversarial manipulation is possible. The minStay(.) function used in Algorithm~\ref{alg:appl_trig}outcomes the minimum valid stay duration (without alarming ADM) at a zone for an occupant given his/her arrival time.
    \end{compactenum}

\begin{table*}[!t]
\centering
\scriptsize
\caption{Case study}
\label{tab:case-studies}
\begin{tabular}{|l|l|cl|l|l|l|l|l|l|l|l|l|}
\hline
\multirow{2}{*}{\textbf{Schedule}}        & \multirow{2}{*}{\textbf{Occupant}} & \multicolumn{1}{l|}{\textbf{Time}}     & \textbf{6:00 PM} & \textbf{6:01 PM} & \textbf{6:02 PM} & \textbf{6:03 PM} & \textbf{6:04 PM} & \textbf{6:05 PM} & \textbf{6:06 PM} & \textbf{6:07 PM} & \textbf{6:08 PM} & \textbf{6:09 PM} \\ \cline{3-13} 
                                          &                                    & \multicolumn{1}{l|}{\textbf{Slot (t)}} & \textbf{1080}    & \textbf{1081}    & \textbf{1082}    & \textbf{1083}    & \textbf{1084}    & \textbf{1085}    & \textbf{1086}    & \textbf{1087}    & \textbf{1088}    & \textbf{1089}    \\ \hline
\multirow{2}{*}{\textbf{Actual}}          & \textbf{Alice}                     & \multicolumn{2}{c|}{2}                                    & 2                & 2                & 2                & 2                & 2                & 2                & 2                & 2                & 2                \\ \cline{2-13} 
                                          & \textbf{Bob}                       & \multicolumn{2}{c|}{0}                                    & 0                & 0                & 0                & 0                & 0                & 0                & 0                & 0                & 0                \\ \hline
\multirow{2}{*}{\textbf{Greedy}}          & \textbf{Alice}                     & \multicolumn{2}{c|}{2}                                    & 2                & 2                & 2                & 2                & 2                & 2                & 2                & 2                & 2                \\ \cline{2-13} 
                                          & \textbf{Bob}                       & \multicolumn{2}{c|}{0}                                    & 0                & 0                & 0                & 0                & 0                & 0                & 0                & 0                & 0                \\ \hline
\multirow{2}{*}{\textbf{SHATTER}}         & \textbf{Alice}                     & \multicolumn{2}{c|}{2}                                    & 2                & 2                & 2                & 2                & 2                & 3                & 3                & 4                & 4                \\ \cline{2-13} 
                                          & \textbf{Bob}                       & \multicolumn{2}{c|}{2}                                    & 2                & 2                & 2                & 2                & 2                & 2                & 2                & 2                & 2                \\ \hline
\multirow{2}{*}{\begin{tabular}[c]{@{}c@{}}\textbf{Range} \\ \textbf{Threshold}\end{tabular}} & \textbf{Alice}                     & \multicolumn{2}{c|}{{[}9 - 30{]}}                         & {[}10 - 27{]}    & {[}10 - 25{]}    & {[}10 - 22{]}    & {[}11 - 16{]}    & {[}11 - 19{]}    & {[}11 - 13{]}    & {[}{]}           & {[}75 - 75{]}    & {[}66 - 75{]}    \\ \cline{2-13} 
                                          & \textbf{Bob}                       & \multicolumn{2}{c|}{{[}5-11{]}}                           & {[}9-18{]}       & {[}5-11{]}       & {[}9-18{]}       & {[}16-19{]}      & {[}9-18{]}       & {[}25-31{]}      & {[}8-18{]}       & {[}9-18{]}       & {[}1-9{]}        \\ \hline
\multirow{2}{*}{\textbf{Trigger Status}}  & \textbf{Alice}                     & \multicolumn{2}{c|}{False}                                & False            & False            & False            & False            & False            & True             & True             & True             & False            \\ \cline{2-13} 
                                          & \textbf{Bob}                       & \multicolumn{2}{c|}{True}                                 & True             & True             & True             & True             & False            & False            & False            & False            & False            \\ \hline
\end{tabular}
\vspace{-9pt}
\end{table*}
\vspace{-6pt}
\begin{algorithm}[hbt!]
\small
\SetAlgoLined
\DontPrintSemicolon
\SetKwFunction{FMain}{ApplianceTriggeringDecision}
\SetKwProg{Fn}{Function}{:}{\KwRet\ }
\Fn{\FMain{$\mathcal{R}$}}{
$trig \leftarrow False$;\\
$arrivalTime \leftarrow 0$;\\
$thresh \leftarrow 0$;\\
$\mathcal{Z} \leftarrow$ set of all zones;\\
\For{$t$ in $Range(\mathcal{T})$}{
\For{$o$ in $Range(\mathcal{O})$}{
$zone \leftarrow \exists_{z \in \mathcal{Z}} \mathcal{\bar{S}}{t,o,z}^{\mathit{OT}}$; \\
\If{$\mathcal{E}{t,o,zone}^{\mathit{A}}$}{
$thresh \leftarrow minStay(t, o, zone)$; \
$arrivalTime \leftarrow t$;
}
\If{$t - arrivalTime \leq thresh$ and $\neg \mathcal{S}_{t,o,zone}^{\mathit{OT}}$}{
$trig \leftarrow True$;
}
}
}
\KwRet $trig$;
}
\normalsize
\caption{Revised Appliance Triggering Decision.}
\end{algorithm}



\section{Case Study}
\label{sec:case_studies}
In this section, we conduct empirical case studies to illustrate the working principle of the SHATTER framework in the case of identifying stealthy attack vectors and corresponding attack costs. To discuss the studies easily, we denote the two occupants of our considered home system as Alice and Bob and the intruder/attacker as Trudy. The occupancy/activity information of the home occupants is taken from the ARAS (Home A) dataset. The case study will be described using Table~\ref{tab:case-studies}. The actual occupancy information in the table is taken from day 4 (6 PM - 6:09 PM). We consider a greedy attack strategy (i.e., demonstrated in Algorithm~\ref{alg:greedy_schedule}.) as the baseline to compare the SHATTER-generated dynamic schedule. In the greedy strategy, we consider that the attacker will schedule the occupant to the zone and activity, which is mapped to the highest cost until the maximum possible stay duration at that particular zone and time.
    %
    \newenvironment{algocolor}{%
   \setlength{\parindent}{0pt}
   \itshape
   \color{emerald}
    }{}

\begin{algorithm}[hbt!]
\label{alg:greedy_schedule}
\small
\SetAlgoLined
\DontPrintSemicolon
\SetKwFunction{FMain}{GreedyScheduleGeneration}
\SetKwProg{Fn}{Function}{:}{\KwRet\ }
\Fn{\FMain{$\mathcal{R}$}}{
$arrivalTime \leftarrow 0$;\\
\While{$arrivalTime < length(\mathcal{T})$}{
\For{$o \in Range(\mathcal{O})$}{
$t \leftarrow arrivalTime$;\\
$zone \leftarrow z \mid \mathcal{\bar{S}}_{t,o,z} \land \mathcal{G}^{\bar{S}}t$ is maximized;
$duration \leftarrow \mathit{maxStay}(t,o,zone)$;\\
\For{$d \in Range(duration)$}{
$\mathcal{\bar{S}}{t,o,z} \leftarrow True$;
}
$arrivalTime \leftarrow arrivalTime + duration$;
}
}
}
\normalsize
\caption{Greedy Schedule Generation.}
\end{algorithm}
    
             

The ARAS zones- Bedroom (Z-1), Livingroom (Z-2), Kitchen (Z-3), and Bathroom (Z-4) incur 0.13\textcent, 0.135\textcent, 2.69\textcent, 0.79\textcent \hspace{0.1cm} respectively for HVAC control of single occupant presence doing the most intensive task in the corresponding zones, while the appliance triggering costs 0.197\textcent, 0.2096\textcent, 1.67\textcent, and 0.83\textcent \hspace{0.1cm} respectively. Hence, the control cost for Alice is 1.6\textcent, while there is no actual benign control cost since Bob was outside the home in all considered 10 slots. The total greedy attack cost for Alice is 1.38\textcent, while the total SHATTER attack cost for her is 10.93\textcent (7.47\textcent for HVAC control and 3.16\textcent for appliance triggering). Hence, the SHATTER attack cost is 8 times than greedy cost for Alice. Similarly, for Bob total SHATTER attack cost is 1.5\textcent. The trigger status in Table~\ref{tab:case-studies} is calculated from Algorithm~\ref{alg:appl_trig}. The primary reason behind the greedy scheduling not performing similarly to the SHATTER attack is that at 5.32 PM, the greedy attack schedule chooses the most rewarding zone (i.e., Kitchen). Consequently, to be consistent with the anomaly detection model, the greedy attack schedule needs to choose the outside zone for Bob, which is the same as his current zone. Since the occupant stays at the exact location as the attack schedule as in Table~\ref{tab:case-studies}, Trudy cannot trigger any other devices in the zones to avoid suspicion of the occupants. On the other hand, SHATTER looks for long-term rewards in a specified time horizon. Accordingly, the SHATTER framework outperforms the greedy approach.

\section{Testbed-Based Validation}
\label{sec:testbed}
%
\begin{figure}[!t]
\vspace{-3pt}
    \begin{center}
        \subfigure[]
        {
        \label{subfig:testbed}
            \includegraphics[width=0.51\columnwidth]{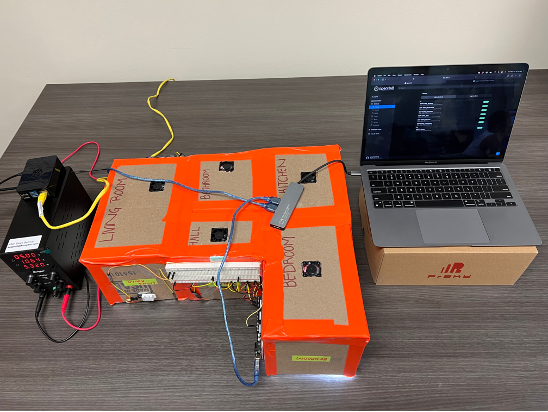}
        }
        \subfigure[]
        {
        \label{subfig:control}
            \includegraphics[width=0.42\columnwidth]{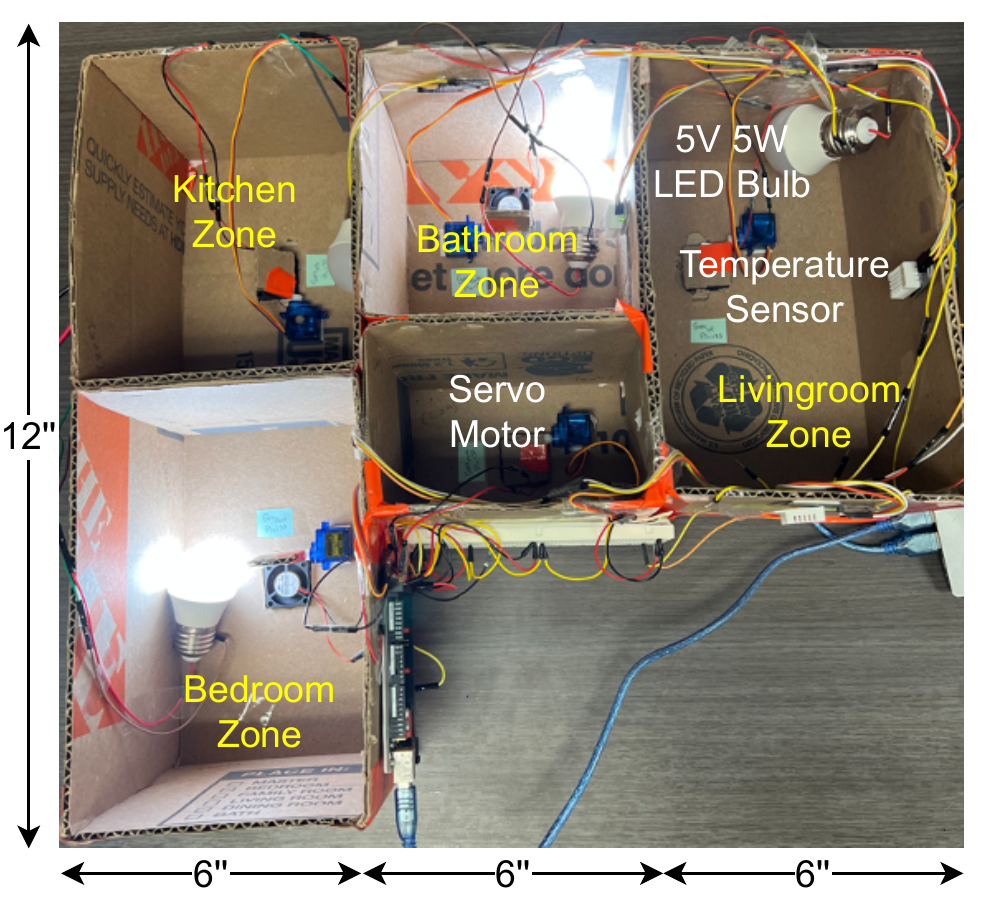}
        }
\end{center}
\vspace{-15pt}
\caption{(a) Demonstrates a testbed instance, when Alice is showering in the bathroom zone, and Bob is taking nap in the Bedroom zone, where (b) shows the benign control scenario, where bathroom and bedroom zone vents are open and is getting air supply to neutralize the added heat generated from the two occupants, corresponding zone lights and the smart bathtub appliance.}
\label{fig:testbed}
\vspace{-20pt}
\end{figure}
%
%
\begin{figure}[t]
\centering
\includegraphics[width = 0.45\textwidth]{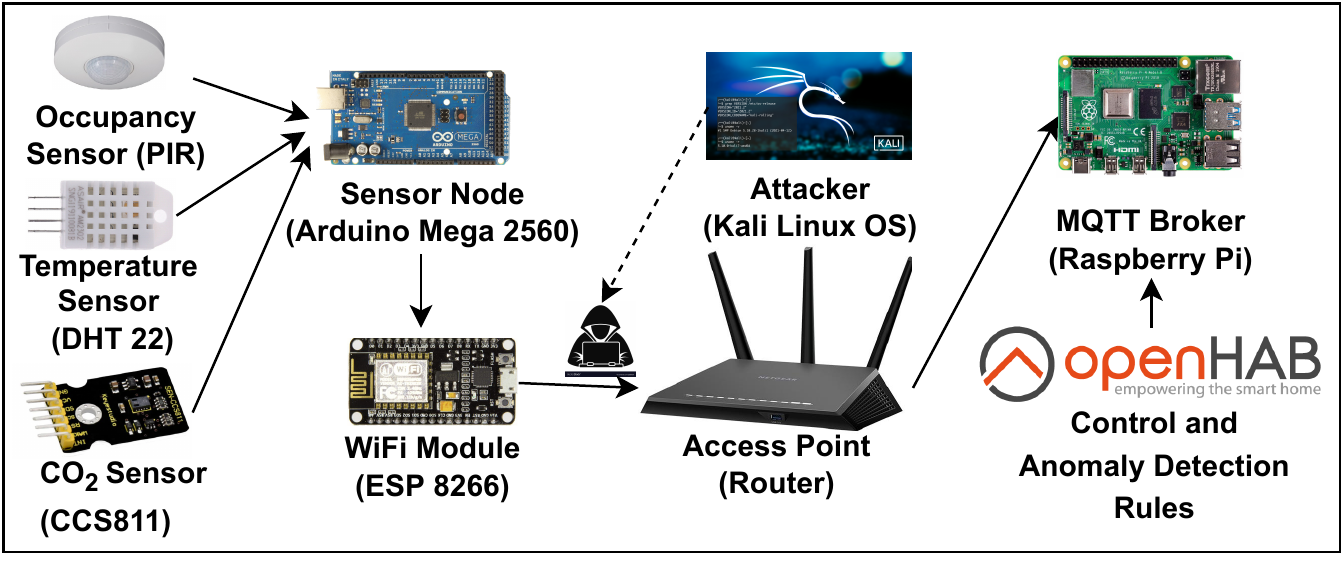}
\centering
\vspace{-9pt}
\caption{Prototype testbed architecture.}
\label{fig:attack_scenarios}
\vspace{-15pt}
\end{figure}
%
We build a prototype testbed for validating our proposed framework. For testbed implementation, we consider that the attacker has full access to measurements and can access the devices. The occupants and appliances are modeled using 5V, 5W led light bulbs in the testbed. The testbed zones are considered from the ARAS testbed. We scaled down the testbed by the scale (=24) in all dimensions. The energy consumption of the occupants and the appliances are scaled accordingly. Our experimentation shows that the temperature and ventilation of the testbed DCHVAC control modeling are not linear. The primary reason behind this is that the zones in the testbed are not completely insulated. For learning the dynamics of the IAQ in the testbed, we trained a regression model for estimating the airflow and heat generation given the temperature. The temperature of the zones is measured using the DHT-22 sensor. We did not consider pollutant generation for the testbed. The emulation of the different activities is carried out by turning on the led lights for a different amount of time. Similarly, the DCHVAC is mimicked by turning on our 1.4 CFM supply fans for a different amount of time identified by the trained polynomial regression model (degree = 2). Such kind of modeling experienced less than 2\% error compared to the testbed measurements. To discuss the studies easily, we denote the two occupants of our considered home system as Alice and Bob and the intruder/attacker as Trudy. Figure~\ref{fig:testbed}(b) shows the benign control situation while at the SHATTER-identified attack scenario. However, in the attacked scenario, the control system gets misinformed that Alice and Bob are cooking without getting alarmed (i.e., ADM is bypassed), so chill air is supplied in the kitchen zone. Eventually, the kitchen zone got more chilled compared to the setpoint letting temperature increment in the occupied zones and energy cost increment. We conducted experimentation by taking 1-hour measurements from the ARAS dataset (house-A). Figure~\ref{fig:attack_scenarios} shows our prototype architecture. We used openHAB as our supervisory control and data acquisition (SCADA) system and attacked the raspberry pi-based MQTT broker to imitate the attack and measure the attack impact. Finally, we found a 78\% increment in energy consumption after the experimentation. Our attack approach here aims at real-time modification of MQTT protocol network messages. To carry out the attack, we employ the Polymorph and the Scapy frameworks. The considered attack is feasible to be launched primarily with a \$35 Raspberry Pi 2 device, which can play the role of a sniffer, MQTT broker, and also packet crafter~\cite{akkaya2015iot, zhu2018tu}.

\section{Evaluation}
\label{sec:evaluation}

This section presents the findings from the considered smart home control model and the feasibility of implementing our proposed framework. We present the SHATTER's evaluation results considering the following set of research questions.

\textbf{RQ1} What is ADM's contribution to reducing stealthy FDI attacks? Section~\ref{subsubsec:anomaly_attack_impact})

\textbf{RQ2} What is the Performance of the SHATTER-generated Attack Schedule? Section~\ref{subsubsec:attack_schedule})

\textbf{RQ3} What is the contribution of an activity monitoring system in the case of aggravating attack impact? (Section~\ref{subsubsec:activity_monitoring_system})

\textbf{RQ4} What are the framework findings in assessing the proposed attack impact with variable attacker's capability? (Section~\ref{subsubsec:attackers_capability})

\textbf{RQ5} How feasible is implementing the proposed framework for a scalable smart home system/other CPS domain? (Section~\ref{subsec:scalability})

\subsection{Evaluation of Anomaly Detection Model}
\label{subsubsec:anomaly_attack_impact}

\begin{table}[t]
\caption{Comparison of ADMs based on the 
attacker's knowledge.}
\label{tab:adm-analysis}
\vspace{-6pt}
\begin{tabular}{|l|l|l|l|l|l|l|}
\hline
ADM                      & \begin{tabular}[c]{@{}l@{}}Attacker's\\ Knowledge\end{tabular}            & Dataset & \begin{tabular}[c]{@{}l@{}}Accu-\\racy\end{tabular} & \begin{tabular}[c]{@{}l@{}}Prec-\\ision\end{tabular} & Recall   & \begin{tabular}[c]{@{}l@{}}F1-\\ Score\end{tabular} \\ \hline
\multirow{8}{*}{DBSCAN}  & \multirow{4}{*}{\begin{tabular}[c]{@{}l@{}}All\\ Data\end{tabular}}     & H1O1    & 0.75     & 0.83      & 0.63     & 0.71                                                \\ \cline{3-7} 
                         &                                                                         & H1O2    & 0.83     & 0.75      & 1.0      & 0.85                                                \\ \cline{3-7} 
                         &                                                                         & H2O1    & 0.73     & 0.71      & 0.76     & 0.73                                                \\ \cline{3-7} 
                         &                                                                         & H2O2    & 0.67     & 0.62      & 0.89     & 0.73                                                \\ \cline{2-7} 
                         & \multirow{4}{*}{\begin{tabular}[c]{@{}l@{}}Partial\\ Data\end{tabular}} & H1O1    & 0.67     & 0.64      & 0.78     & 0.70                                                \\ \cline{3-7} 
                         &                                                                         & H1O2    & 0.63     & 0.57     & 1.0      & 0.73                                                \\ \cline{3-7} 
                         &                                                                         & H2O1    & 0.61     & 0.56      & 0.70 & 0.38                                                \\ \cline{3-7} 
                         &                                                                         & H2O2    & 0.56     & 0.54     & 0.89 & 0.67                                                \\ \hline
\multirow{8}{*}{\begin{tabular}[c]{@{}l@{}}K-Means\\ Clustering\end{tabular}} & \multirow{4}{*}{\begin{tabular}[c]{@{}l@{}}All\\ Data\end{tabular}}     & H1O1    & 0.71     & 0.77      & 0.6     & 0.67                                                \\ \cline{3-7} 
                         &                                                                         & H1O2    & 0.93     & 0.87      & 1        & 0.93                                                \\ \cline{3-7} 
                         &                                                                         & H2O1    & 0.76     & 0.8      & 0.69     & 0.74                                                \\ \cline{3-7} 
                         &                                                                         & H2O2    & 0.88     & 0.82      & 0.96     & 0.88                                                \\ \cline{2-7} 
                         & \multirow{4}{*}{\begin{tabular}[c]{@{}l@{}}Partial\\ Data\end{tabular}} & H1O1    & 0.64     & 0.6      & 0.87     & 0.70                                                \\ \cline{3-7} 
                         &                                                                         & H1O2    & 0.71     & 0.62      & 1.0      & 0.77                                               \\ \cline{3-7} 
                         &                                                                         & H2O1    & 0.60    & 0.56      & 0.88     & 0.68                                                \\ \cline{3-7} 
                         &                                                                         & H2O2    & 0.67     & 0.62      & 0.89     & 0.73                                                \\ \hline
\end{tabular}
\vspace{-12pt}
\end{table}

As we discussed, stealthy FDI attacks can help knowledgeable adversaries to bypass the ADM and make the system vulnerable. Hence, assessing the system's ADM against stealthy FDI attacks is mandatory from a security perspective. Here, we compare the SHATTER-considered ADM to a state-of-the-art framework (i.e., BIoTA) that does not consider a robust ADM but rather considers some verification rules. The contribution of ADM is evaluated based on the reduction of stealthy FDI attack impact. In this attack impact evaluation process, we do not consider the triggering of the smart appliances. A comprehensive performance evaluation of the considered ADMs is provided in Table~\ref{tab:adm-analysis}. The results indicate that other than the H1O1 dataset, K-Means clustering outperformed DBSCAN-based ADM. We generate the anomalous/attack data using the BIoTA framework for this evaluation process. We consider variable attackers' knowledge ADM assessment, i.e., the attacker has either access to all day's occupancy, activity, and sensor measurement data used for ADM training (all data) or 50\% of them (partial data).

Table~\ref{tab:adm_contribution} shows the cost comparison between the BIoTA, greedy attack scheduling, and our proposed SHATTER framework for both ARAS Houses A and B datasets. BIoTA-identified attack vectors' costs are at most 1.5 times higher than the SHATTER-identified attack vectors. However, our considered ADM identified (60-100)\% attack vectors identified by the BIoTA framework as anomalies. Hence, we can conclude that the BIoTA-identified attack vectors are not stealthy for the considered ADMs. In our proposed framework, we consider a robust ADM to synthesize critical and hazardous attack vectors that can evade modern control systems and thus obtain a defense guide for secure control architecture. 

There is an interesting insight from the results shown in Table~\ref{tab:adm_contribution}. It seems that the attack impact of DBSCAN-based ADM is lower than the K-Means clustering-based ADM although the latter mostly showed better performance (Table~\ref{tab:adm-analysis}). The attack impact of the DBSCAN-based system can be as much as 35\% lower than that of the K-means clustering-based ADM. The reason behind this is that the attacks obtained from BIoTA were very naive and maintained a large margin from the benign data distribution. Hence, the BIoTA-identified attacks are not a good choice for ADM model assessment. Since K-means clustering clusters all the training samples, unlike DBSCAN, which removes the noise points, the cluster areas were unnecessarily large. Accordingly, the k-means cluster-based ADM failed to capture the zero-day attacks found by SHATTER. From this evaluation, it can be inferred that the SHATTER-identified attack vectors are more appropriate to assess ADMs than the state-of-the-art. It needs to be noted that the purpose of this work is not to propose an optimal ADM. The developed ADM is used for the experimentation and evaluation of SHATTER. The framework is flexible enough to consider any ADM to assess its data-driven security and robustness against stealthy FDI/ integrity attacks in IoT-based control systems. In the rest of the evaluation, we will use DBSCAN-based ADM as it performs better.

\begin{table}[]
\small
\centering
\caption{Comparison in between SHATTER Attack Impact with BIoTA framework and Greedy Attack Scheduling Approach.}
\vspace{-6pt}
\scriptsize
\label{tab:adm_contribution}
\begin{tabular}{|l|l|l|l|l|}
\hline
\textbf{\begin{tabular}[c]{@{}l@{}}Framework/\\ Approach\end{tabular}} & \textbf{ADM}                                                                  & \textbf{\begin{tabular}[c]{@{}l@{}}Attacker's\\ Knowledge\end{tabular}} & \textbf{\begin{tabular}[c]{@{}l@{}}House A\\ Energy\\ Cost (\$)\end{tabular}} & \textbf{\begin{tabular}[c]{@{}l@{}}House B\\ Energy \\ Cost (\$)\end{tabular}} \\ \hline
BIoTA                                                                  & Rules-based                         & \multicolumn{1}{c|}{-}                                                                                      & 775.83                                                                        & 518.50                                                                         \\ \hline
\multirow{4}{*}{Greedy}                                                & \multirow{2}{*}{DBSCAN}                                                       &  All Data                                                                                                        & 517.51                                                                        & 307.06                                                                         \\ \cline{3-5}
                                                                       &                                                                               &  Partial Data                                                                                                     & 447.02                                                                        & 148.39                                                                         \\ \cline{2-5} 
                                                                       & \multirow{2}{*}{\begin{tabular}[c]{@{}l@{}}K-Means\\ Clustering\end{tabular}} & All Data                                                                                                        & 513.92                                                                        & 220.37                                                                         \\ \cline{3-5} 
                                                                       &                                                                               & Partial Data                                                                                                     & 30.4.90                                                                       & 104.61                                                                         \\ \hline
\multirow{4}{*}{SHATTER}                                               & \multirow{2}{*}{DBSCAN}                                                       &  All Data                                                                                                        & 549.58                                                                        & 299.69                                                                         \\ \cline{3-5} 
                                                                       &                                                                               &  Partial Data                                                                                                  & 461.01                                                                        & 132.95                                                                         \\ \cline{2-5} 
                                                                       & \multirow{2}{*}{\begin{tabular}[c]{@{}l@{}}K-Means\\ Clustering\end{tabular}} & All Data                                                                                                      & 745.04                                                                        & 454.61                                                                         \\ \cline{3-5} 
                                                                       &                                                                               &  Partial Data                                                                                                   & 361.15                                                                        & 434.09                                                                         \\ \hline
\end{tabular}
\vspace{-9pt}
\end{table}

\vspace{-3pt}
\subsection{Evaluation of SHATTER-generated Attack Schedule}
\label{subsubsec:attack_schedule}
The SHATTER framework generates an optimal attack schedule to misinform the controller with occupancy information at different zones. As discussed before, the considered scheduling is an NP-hard problem. To get a polynomial time solution, we consider window-based dynamic optimization. For the experimentation, we optimize each at every 10 slots (a slot corresponds to 1 minute sampling time) to get the final attack schedule. However, throughout the day, there can be 1440 possible slots (considering the sampling time to be 1 minute). Reducing sampling time will make the scheduling problem even more critical, in turn, a robust control system. Our experimentation suggests that a SHATTER-generated attack schedule can incur significantly higher costs than a greedy attack scheduling strategy. From Table~\ref{tab:adm_contribution}, we can see that the proposed attack scheduling is incurring up to \$32.07 more cost (i.e., for DBSCAN-based ADM of House A) throughout a month compared to the greedy attack approach, where the benign control cost was \$244.69. The higher attack cost for the greedy attack schedule compared to the SHATTER attack schedule supports the essence of dynamic attack scheduling. The SHATTER-identified attack vector would create more impact if the optimization window was larger.

\vspace{-3pt}
\subsection{Evaluation of Activity Monitoring System}
\label{subsubsec:activity_monitoring_system}
\begin{figure}[!t]
    \begin{center}
         \subfigure[]
        {
        \label{subfig:activity_monitoring_house_A}
            \includegraphics[width=0.46\columnwidth]{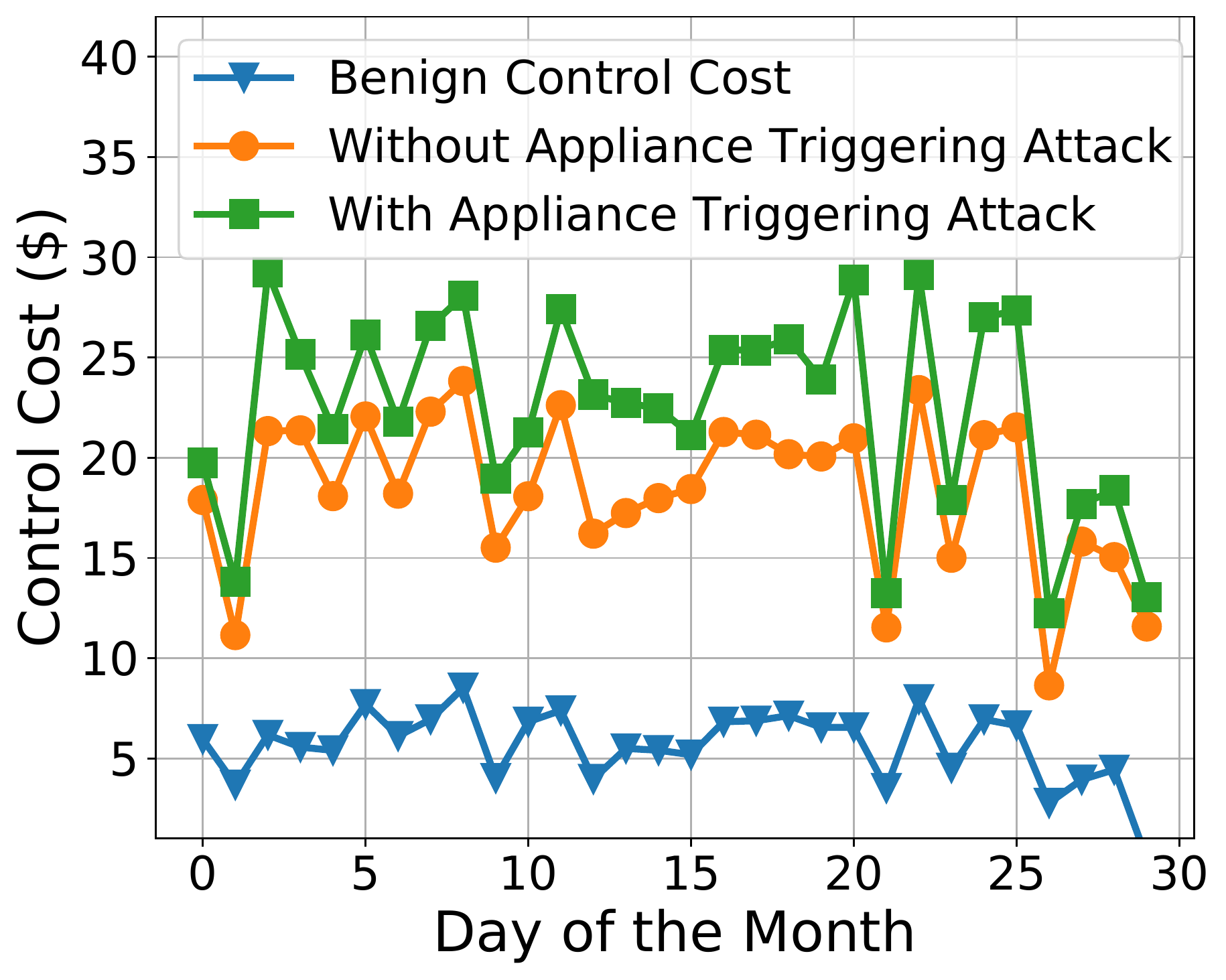}
        }
        \subfigure[]
        {
        \label{subfig:activity_monitoring_house_A}
            \includegraphics[width=0.46\columnwidth]{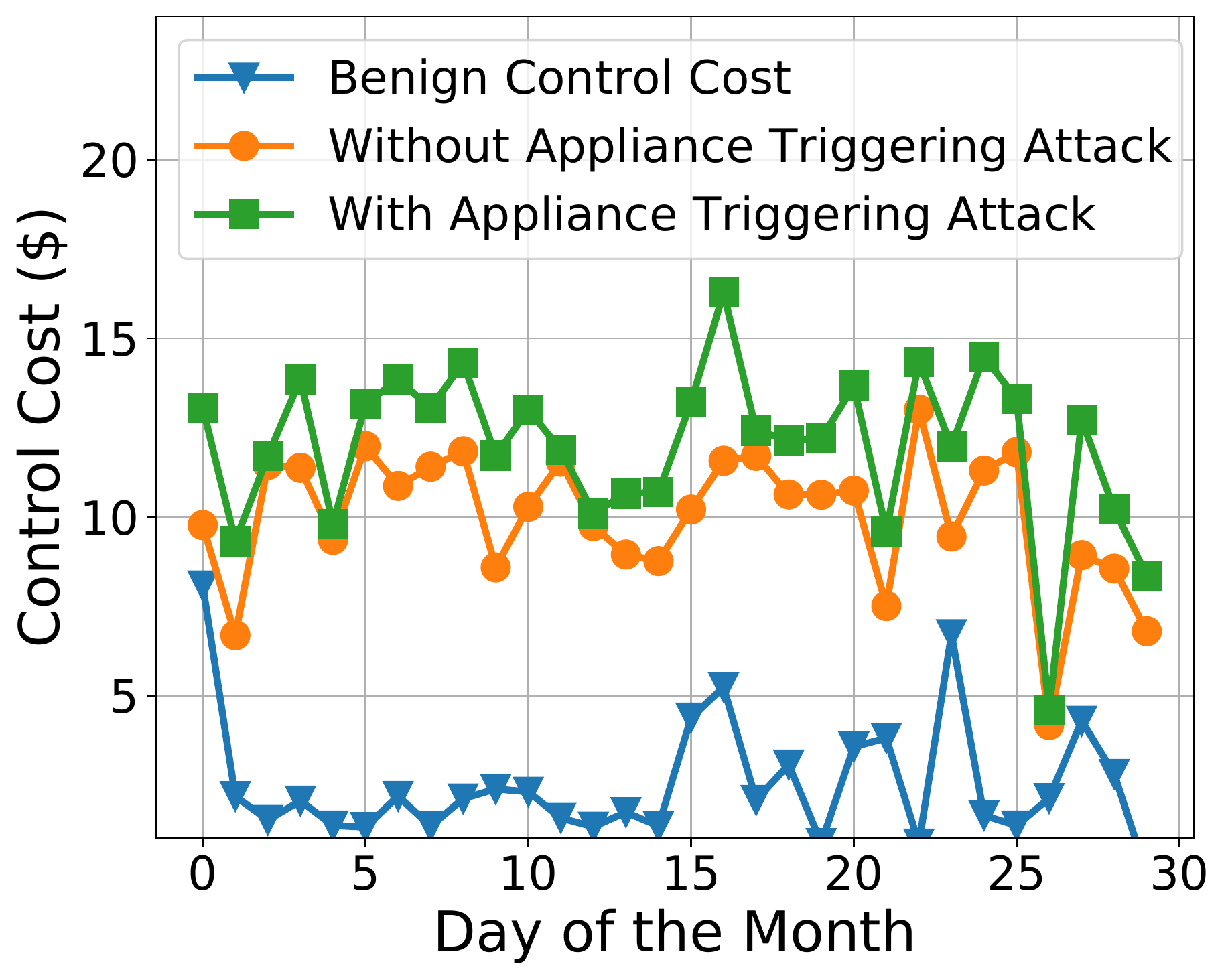}
        }     
    \end{center}
    \vspace{-14pt}
    \caption{Control cost comparison with or without appliance triggering attacks for (a) ARAS House A (b) ARAS House B considering DBSCAN ADM.}
    \label{fig:activity_monitoring}
    \vspace{-14pt}
\end{figure}
In this evaluation, we show how the knowledge of activity information can help the adversaries to further increase the smart home HVAC energy cost. With the knowledge of activity information, a stealthy appliance-triggering attack is possible. Without the activity information, an attacker can maliciously activate an appliances that can create suspicion among the occupants even after evading the ADM. In the earlier evaluations, we did not consider any appliance-triggering attacks. The spikes in Figure~\ref{fig:activity_monitoring} indicate a significant rise in control cost through appliance-triggering attacks. With full adversarial access, such an attack can increase the cost by \$123.09 (+22.73\%) and \$60.03 (+20.03\%), respectively, for ARAS Houses A and~B.

\vspace{-3pt}
\subsection{Attack Impact Evaluation by Varying Attacker's Capability}
\label{subsubsec:attackers_capability}
We evaluate the attack impact with different attackers' capabilities. In this evaluation, we analyze the appliance-triggering attack impact with variable measurement and appliance access. In Table~\ref{tab:ams_contribution_zones}, we show the attack impact considering that the adversary has accessibility to sensor measurements of different zones and can trigger all appliances. The evaluation shows that the attacker can create a significant attack impact by having access to 3 and 4 zones. However, access to 2 zones reduces the attack impact drastically (3.7 times in ARAS House A and 12.22 times in House B). Hence, SHATTER proposes a successful defense strategy. In Table~\ref{tab:ams_contribution_appliance}, we show the attack impact considering that the adversary has accessibility to appliances of different zones and can inject false measurements in all zone sensor measurements. The results show that even with access to 3 appliances (out of 13), a significant attack impact can be created in both houses. The analysis from Tables~\ref{tab:ams_contribution_zones} and~\ref{tab:ams_contribution_appliance} suggests that the defense mechanism should focus on securing occupancy and IAQ measurements compared to appliances.

\begin{table}[]
\small
\centering
\caption{Appliance Triggering Attack Impact with Various Zone Measurement Access Capability}
\vspace{-6pt}
\scriptsize
\label{tab:ams_contribution_zones}
\begin{tabular}{|l|l|l|}
\hline
\multicolumn{1}{|l|}{\begin{tabular}[c]{@{}l@{}}\textbf{Number of } \\ \textbf{Accessible Zone}\end{tabular}}                                  & \begin{tabular}[c]{@{}l@{}}\textbf{House A} \\ \textbf{Energy Cost (\$)}\end{tabular} & \begin{tabular}[c]{@{}l@{}}\textbf{House B} \\ \textbf{Energy Cost (\$)}\end{tabular} \\ \hline
4 Zones                                                               & 124.93           & 60.03          \\ \hline
\begin{tabular}[c]{@{}l@{}}3 Zones\end{tabular}  & 117.42           & 31.91          \\ \hline
\begin{tabular}[c]{@{}l@{}}2 Zones\end{tabular} & 33.74           & 4.91           \\ \hline
\end{tabular}
\vspace{-9pt}
\end{table}

\begin{table}[t]
\small
\centering
\caption{Appliance Triggering Attack Impact with Various Appliance Triggering Capability}
\label{tab:ams_contribution_appliance}
\vspace{-6pt}
\scriptsize
\begin{tabular}{|l|l|l|}
\hline
\multicolumn{1}{|l|}{\begin{tabular}[c]{@{}l@{}}\textbf{Number of } \\ \textbf{Accessible Appliances}\end{tabular}}                                  & \begin{tabular}[c]{@{}l@{}}\textbf{House A} \\ \textbf{Energy Cost (\$)}\end{tabular} & \begin{tabular}[c]{@{}l@{}}\textbf{House B} \\ \textbf{Energy Cost (\$)}\end{tabular} \\ \hline
13 Appliances                                                              & 124.93           & 60.03           \\ \hline
\begin{tabular}[c]{@{}l@{}}8 Appliances\end{tabular}  & 117.89           & 51.16        \\ \hline
\begin{tabular}[c]{@{}l@{}}3 Appliances\end{tabular} & 93.05          & 50.82          \\ \hline
\end{tabular}
\vspace{-9pt}
\end{table}

\vspace{-3pt}
\subsection{Scalability Analysis of SHATTER}
\label{subsec:scalability}
\begin{figure}[!t]
    \begin{center}
         \subfigure[]
        {
        \label{subfig:scalability_time_horizon}
            \includegraphics[width=0.48\columnwidth]{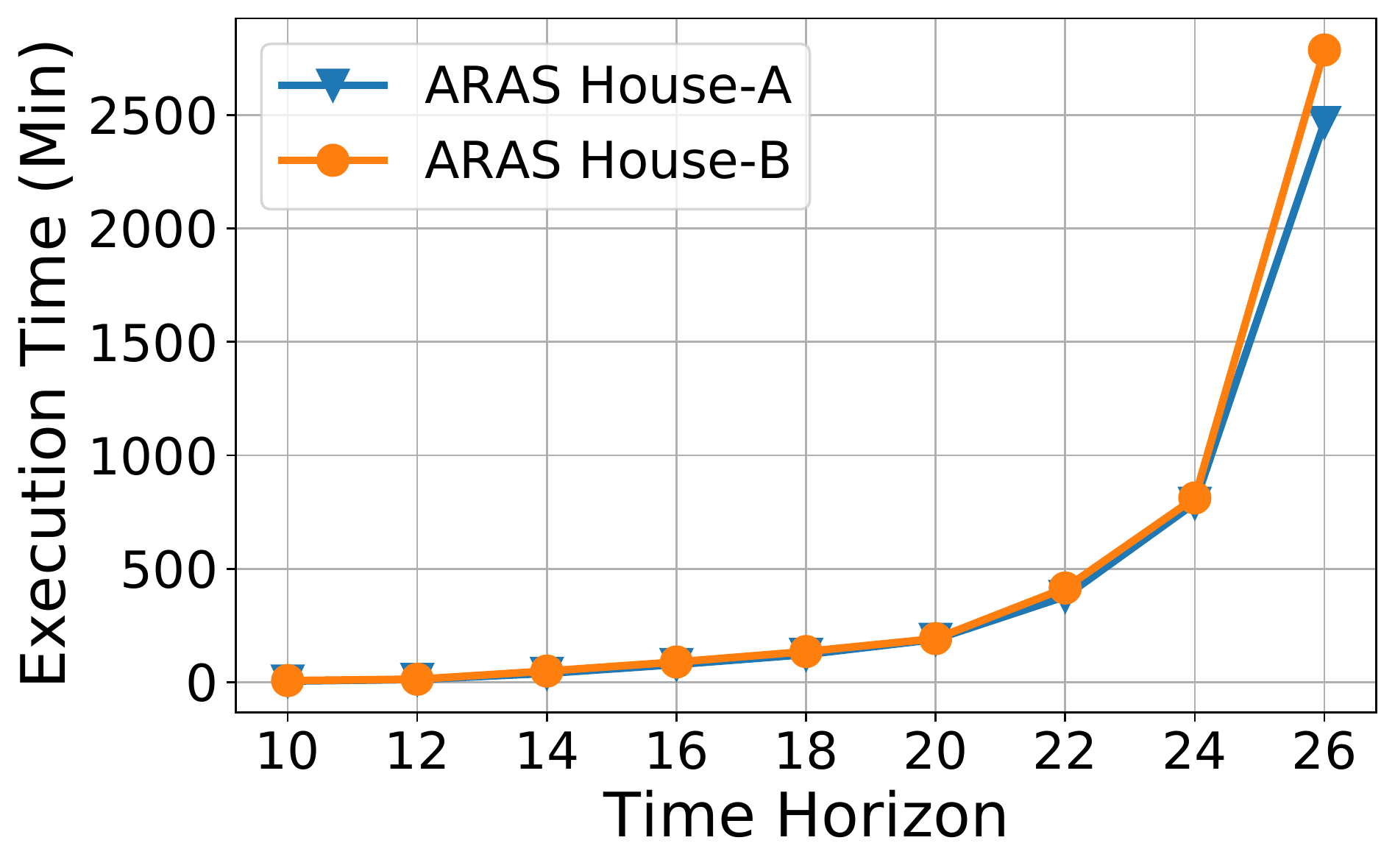}
        }
        \hspace{-8pt}
        \subfigure[]
        {
        \label{subfig:scalability_horizontal}
            \includegraphics[width=0.46\columnwidth]{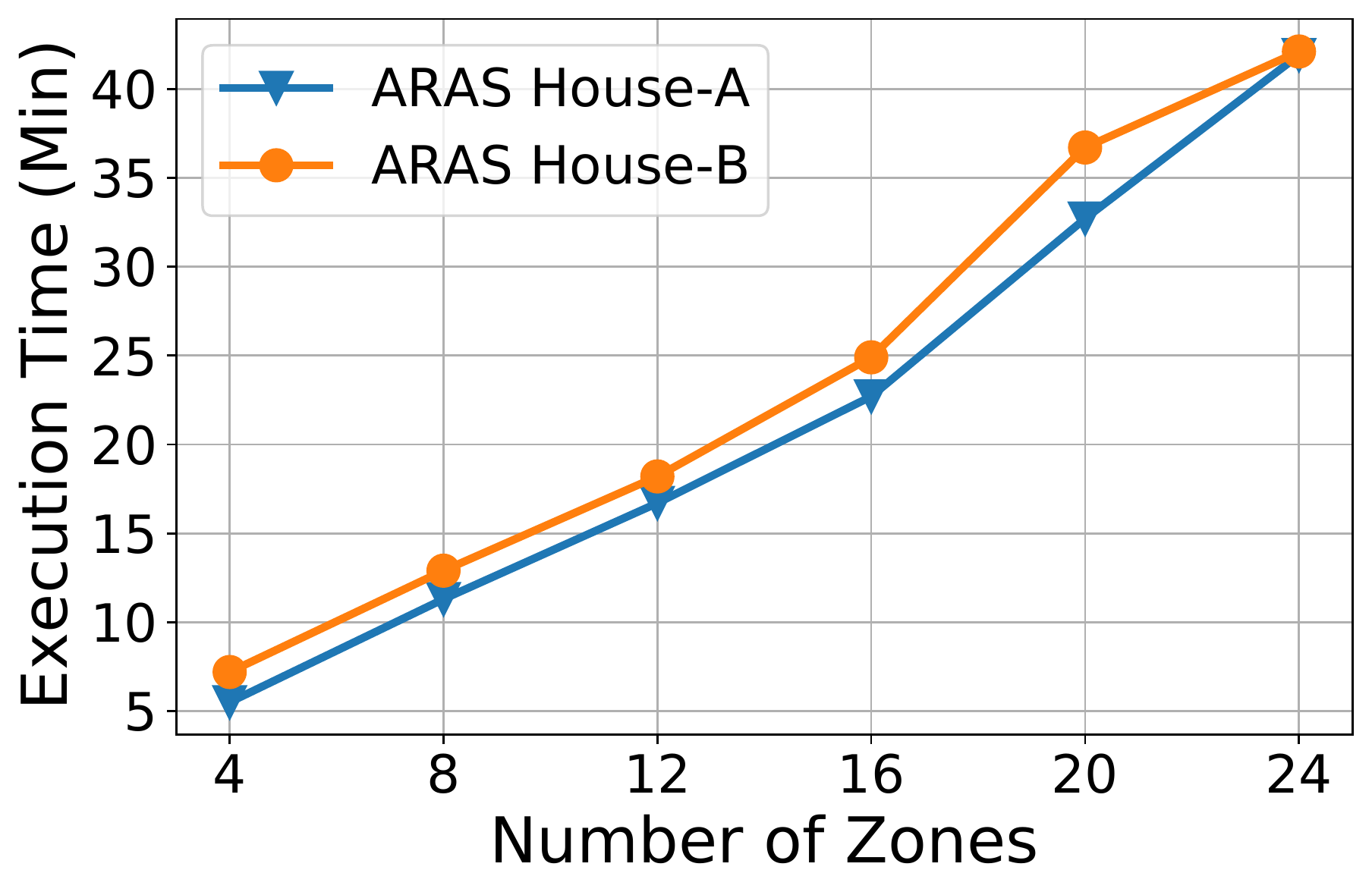}
        }     
    \end{center}
    \vspace{-12pt}
    \caption{Scalability analysis based on (a) time horizon, (b) horizontal scaling.}
    \label{fig:scalability}
    \vspace{-15pt}
\end{figure}

Identification of an optimal stealthy attack vector through the proposed attack technique evading the complex ADM is an NP-hard problem. The optimal attack vector identification requires solving an optimization problem of 1440 (1 minute sampling time for all measurements) lookback time. We evaluate the scalability of the SHATTER framework by varying the time horizon (i.e., lookback time). The increase in the number of zones multiplies the number of constraints. Fig.~\ref{subfig:scalability_time_horizon} shows scalability analysis based on different lookback times, which affirms the exponential growth of execution time. However, horizontal scaling (i.e., increased number of zones) raises the number of constraints linearly. Hence, the execution time for an increased number of zone is growing linearly, as shown in Figure~\ref{subfig:scalability_horizontal} (lookback = 10).


\section{Related Work}
\label{sec:related-works}
In this section, we compare our proposed attack analyzer with comparable literature. Although the work mainly focuses on attack analysis, we provide a comprehensive literature review of the control systems, ADM, and attack analytics.

\subsection{Control Systems and Anomaly Detection}
Traditionally, closed-loop control systems have relied on physics-based models of dynamic systems for optimal control decisions. These models are mathematically analyzed. However, in recent times, ML is increasingly used to develop controllers. Since ML models are integrated into controllers as a classifier to identify control actions or as a validator to detect faulty/anomalous/attacked data~\cite{moghaddamenergy, peng2018using, elkhoukhi2018towards}. Although ML-based methods are adaptable and robust against measurement errors and noises, they lack systematic mathematical analysis and are often viewed as black-box methods. 

There has been extensive research into developing ADMs, in the context of smart homes and related domains. For instance, Pan et al. introduced an ADM that utilizes a context-aware BACnet data structure for building automation and control networks~\cite{pan2019context}. However, the ADM did not prove effective in reducing false-positive alarms. Another research developed a lightweight rule-based ADM for smart home/building control systems~\cite{luo2013rule}. Analyzing BIoTA, we have discovered that rules-based ADMs leave backdoors to be exploited, which makes them vulnerable to zero-day attacks. In one of our previous works, we developed an ensembled unsupervised ML model for detecting attacks from the BIoTA framework~\cite{haque2021ensemble}. The performance was extensively good although only getting trained with benign data, we have already experimented that most of the BIoTA-identified attacks are trivial. The definition of performance from different ADM-related research is questionable. The reason is that, in most cases, the performance of the models is evaluated based on some known attack or concrete decision boundary around the known benign samples. Hence, the vulnerability of those models can be easily exploited. Therefore, we focus on critical feature selection for model training.

\subsection{Attack Analytics}
Development of attack analytics has always been a concern for safety-critical CPSs. The existing research can be broadly classified into regulation, rules, and ML-based analytics.

\noindent \textbf{Regulation and Rules-based Analytics:} Stellios et al. introduced a new approach to identify and evaluate attack paths against critical IoT systems based on assessing risks, utilizing existing tools (i.e., CVE, CVSS)~\cite{stellios2021assessing}. Akatyev et al. evaluated potential threats for futuristic smart homes with multiple diverse components and advanced decision-making abilities~\cite{akatyev2019evidence}. Casola et al. proposed an automated method for threat modeling and risk assessment based on a threat catalog created during the FP7 SPECS project, which targeted the communication protocol and software elements of IoT systems~\cite{casola2019toward}. The analysis of security threats and resiliency in rule-based IoT systems has been thoroughly researched and explored in previous studies. Mohsin et al. developed a formal security analysis framework for IoT-based systems by analyzing network topology and interdependencies between system components~\cite{mohsin2016iotsat, mohsin2017iotchecker}. This framework can identify potential attack vectors from integrity and availability types of attacks and assess the system's resiliency against attackers with varying accessibility and capabilities. However, these proposed frameworks are limited to analyzing the security of rules-based IoT systems. Analysis of such systems does not require investigating historical data or maintaining time-series patterns. SHATTER can find hazardous attack vectors that can evade sophisticated defense tools.

\noindent \textbf{ML-based Analytics:} Solving constraints in ML-based systems is much more complex than in rule-based systems, and as a result, formal analysis of deep neural network-based ML models has become a focus of contemporary research. Various effective tools, including Reluplex, Sherlock, and Marabou, have been developed for verifying ML models~\cite{katz2017reluplex, dutta2017output, katz2019marabou}. Researchers have attempted to identify issues and analyze the behavior of ML-based systems in uncertain environments using formal methods. Souri et al. formally verified a hybrid ML-based approach for fault prediction in IoT applications~\cite{souri2020formal}. However, unlike these verification approaches, the proposed framework can synthesize attack vectors contemplating activity recognition, appliance triggering, and ML-based ADMs, producing useful attack vectors to assess and propose defense systems.


\section{Conclusion}
\label{sec:conclusion}
In this work, we propose a novel framework that analyzes the threat space of a smart IoT-enabled home control system, efficiently extracting ADM rules. We evaluate the proposed attack analyzer's effectiveness on the ARAS dataset. Moreover, we also build a prototype testbed for validating the framework in real-life settings. Experimental analysis using the verification dataset and validation testbed exhibits the effectiveness of the proposed framework. SHATTER generates sub-optimal attack vectors by creating optimal attack schedules in constrained time horizons. The results show that SHATTER-generated attack vectors can increase a home's energy consumption by more than 20\% by leveraging appliance-triggered attacks. Some modern homes generate (i.e., using generators or renewable energy sources) and store energy, using batteries to reduce peak hour energy expense. Based on the capacity of the energy storage, excess energy can be produced, which has not been modeled by our proposed framework. If there is excess energy production, the home can be viewed as a microgrid, which sells the excess energy to the grid. Even though SHATTER-identified attacks will unquestionably decrease earnings compared to a benign operating condition, the attack's impact in this scenario will be distinct and needs attention. In our future attack modeling, we will factor in renewable power sources.

\section{Acknowledgment}\label{sec:acknowledgment}
%
This work is partially supported by the National Security Agency (NSA) under Award\# H98230-21-1-0324, the Department of Energy (DOE) under Award\# DE-CR0000024, and the Visiting Faculty Research Program (VFRP) with the Information Assurance Branch of the AFRL/RI, Rome, NY, and the Information Institute (II). Any opinions, findings, conclusions, or recommendations expressed in this material are those of the authors and do not necessarily reflect the views of the DOD/AFRL, NSA, or DOE.

\bibliographystyle{unsrt}
\bibliography{References}

\end{document}